\RequirePackage{fix-cm}
\documentclass[smallextended]{svjour3}       
\smartqed  
\usepackage{graphicx}
\usepackage[utf8]{inputenc}
\usepackage{amssymb}
\usepackage{amsmath}
\usepackage{theorem}
\usepackage{latexsym}
\usepackage{array}
\usepackage{comment}
\theoremstyle{break}

\newtheorem{Proposition}{Proposition}
\newtheorem{Lemma}{Lemma}
\newtheorem{Corollary}{Corollary}
\newcommand{\DIS}{\displaystyle}

\newcommand{\e}{\mbox{e}}

\newcommand{\Z}{\mathbb{Z}}

\newcommand{\R}{\mathbb{R}}

\newcommand{\Proof}{\hfil\break{\bf Proof}\;\;\;\;}

\begin{document}

\title{Pattern formation of vascular network in a mathematical model of angiogenesis
}


\author{Jun Mada 
 \and
        Tetsuji Tokihiro 
}


\institute{Jun Mada\at
              College of Industrial Technology, 
  Nihon University, 2-11-1 Shin-ei, Narashino, Chiba 275-8576, Japan  \\
             Tel.: +81-47-474-2828\\
              Fax: +81-47-473-2950\\
              \email{mada.jun@nihon-u.ac.jp}           
           \and
          Tetsuji Tokihiro \at
            Graduate School of Mathematical Sciences, the University of Tokyo, 3-8-1 Komaba, Meguro-ku, Tokyo 153-8914, JAPAN   \\
              Tel.: +81-3-5465-7080\\
              Fax: +81-3-5465-7011\\
              \email{toki@ms.u-tokyo.ac.jp}    
}

\date{Received: date / Accepted: date}

\maketitle

\begin{abstract}
We discuss the characteristics of the patterns of the vascular networks in a mathematical model for angiogenesis.
Based on recent in vitro experiments, this mathematical model assumes that the elongation and bifurcation of blood vessels during angiogenesis are determined by the density of endothelial cells at the tip of the vascular network, and describes the dynamical changes in vascular network formation using a system of simultaneous ordinary differential equations. The pattern of formation strongly depends on the supply rate of endothelial cells by cell division, the branching angle, and also on the connectivity of vessels. By introducing reconnection of blood vessels, the statistical distribution of the size of islands in the network is discussed with respect to bifurcation angles and elongation factor distributions. The characteristics of the obtained patterns are analysed using multifractal dimension and other techniques.
\end{abstract}

\section{Introduction}
\label{sec1}
Angiogenesis is a phenomenon in which a new network of blood vessels is formed from an existing network of blood vessels\cite{Risau,CLF}.
Since angiogenesis plays an essential role in the growth and metastasis of cancer cells, as well as in the recovery of blood vessels lost due to injury, etc., elucidating the mechanism of angiogenesis is an important issue in medical sciences\cite{Kerbel,ZBTZX}.
In angiogenesis, vascular endothelial cells (ECs) proliferate due to vascular endothelial growth factor (VEGF) and other factors, vascular wall cells and basement membranes are degraded, ECs migrate to the outside of blood vessels, and a new vascular network is formed.
A lot of models for angiogenesis have been and is being constructed with various mathematical methods from different point of view\cite{AC,TongYuan,Gambaetal,BJJ,SciannaPreziosi,SADP,DaubMerks,Sugiharaetal}.
In recent in vitro experiments, the dynamics of ECs were observed by time-lapse imaging in detail, and it was found that the ECs exhibit very complex phenomena so called cell-mixing  (tip cells are constantly replaced by fast-moving followers that overtake other ECs)\cite{Arima-Nishiyama}. 
Based on these results, Matsuya, Yura, Kurihara and the authors proposed a mathematical model of angiogenesis using a discrete dynamical system that reproduces the elongation and bifurcation of blood vessels\cite{MYMKT}, and then approximated this model with a continuous system of differential equations that introduces the effects of cell division and VEGF\cite{MMYKT}. 
This continuous model qualitatively discusses influence of cell division and VEGF on the vascular network patterns, but it does not include effects of reconnection of blood vessels after branching.
Hence the pattern, thus formed, is topologically equivalent to a Cayley tree. 
When the vascular branches extend in three dimensions, reconnection of blood vessels will occur only rarely, but blood vessels in a retina and those in in vitro experiments in petri dishes are almost confined to a two-dimensional plane.
In these cases, we have to consider reconnection of blood vessels in actual vascular networks.
\begin{figure}[htbp]
\begin{center}
\includegraphics[scale=0.5]{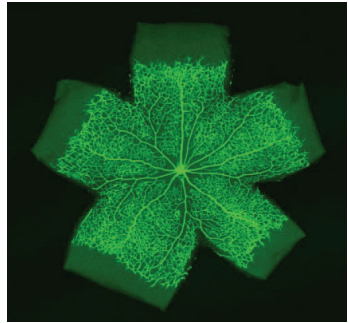}
\caption{Retinal vascular network. (Nature Japan, Nature digest, Vol.12, No.2, Japanese Author. 
https://www.natureasia.com/ja-jp/ndigest/v12/n2/)}
\label{fig:retina}
\end{center}
\end{figure}
A vascular network, retinal blood vessels in particular, has been considered to be realization of fractal growth in finite size\cite{FMP,Masters}.   
Fractal geometry is an important tool to analyse complex objects whose geometry cannot be characterized by an integral dimension\cite{Mandelbrot}.   
Irregular or fragmented shape appeared in nature is sometimes unable to be analysed by traditional Euclidean geometry.
Complex phenomena often exhibit spatial or temporal scaling laws and power-law behaviour and characterized by a measure called fractal dimension\cite{Mandelbrot2,Falconer}.
If an object has self-similarity, its fractal dimension (or self-similarity dimension) is evaluated precisely, however, most complicated objects have various kinds of scaling components and we need notion of multifractal\cite{HJKPS,LopesBetrouni}. 
Since a vascular network in retina usually does not have self-similarity,  multifractal methods are applied to elucidate its characteristics\cite{StosicStosic,SMA}.

In the present paper, we reformulate the previous continuous model\cite{MMYKT} including reconnection of blood vessels.
We are particularly interested in the various patterns of the vascular network. 
Considering reconnection, the angle of branching as well as the ratio of successive branches is very important for pattern formation.
We investigate the dependence of the features of these patterns on the angle and the ratio in detail.
We show the statistical features of the length of blood vessels at each bifurcation, those of the size of an island (the region surrounded by connected blood vessels), total area of the vascular networks and so on.
Fractal and multifractal dimensions of the vascular patterns in the present model are also obtained.
In section~\ref{chap:Math_model}, we explain the model which is described with simultaneous differential equations of time evolution of branch lengths. Several analytically solvable cases are listed. In section~\ref{sec:R-network}, effects of reconnection of blood vessels are investigated. The dependence of bifurcation angles and deviation of parameters on the distribution of island area, length of branches, total area of the network are discussed. 
The fractal nature of the vascular networks obtained in the present mathematical model is discussed in section~\ref{Chapter:fractal}. The conditions for no reconnection and generation of self-similar patterns are also given. 
Section~\ref{Chap:conclusion} is devoted to concluding remarks.

\section{A mathematical model for elongation and bifurcation of blood vessels } 
\label{chap:Math_model}

In the previous paper\cite{MMYKT}, we proposed a phenomenological model for angiogenesis described by simultaneous ordinary differential equations.  The model is based on the facts observed in the in vitro experiments\cite{Arima-Nishiyama};
\begin{itemize}
\item Elongation is caused by a tip cell which is pushed forward by ECs behind it.
\item A vessel divides into not more than two branches at a junction.
\item Bifurcation occurs mostly at the tip when the tip is crowded by ECs.
\item A vessel splits into two branches with an angle of approximately 60 degree.
\end{itemize}
From these observations, we constructed a mathematical model for time evolution of a vascular network by assuming that the average number of ECs near the tip determines elongation and bifurcation of blood vessels.
For a growing branch at the tip, let $L(t)$ be its length at time $t$ and it satisfies
\[
\frac{d L}{d t}(t)=\tilde{F}(n(t)),
\]
where $n(t)$ is the average number of ECs near the tip of the branch, and $\tilde{F}$ is a function which determines elongation and bifurcation of the branch.
A branch elongates if $n(t)$ is larger than a threshold value $n_e$, and it bifurcates when $n(t)$ reaches the threshold value $n_b$. 
We adopted the form of $\tilde{F}(n)$ as
\[
\tilde{F}(n)=\left\{
\begin{array}{cl}
\lambda(n-n_e)&\quad (n_e \le n \le n_b)\\
0& \quad (\mbox{otherwise})
\end{array}
\right..
\]
The parameter $\lambda$ depends on the spatial position of the branch and represents the effects of VEGF and extracellular matrix. 
To determine $n(t)$, we assumed that the density of ECs in the network is spatially uniform and supply rate of ECs is given by a logistic type function. 
By numerical simulation, we solved the simultaneous equations of all the growing branches and examined temporal change of the network, the length distribution of the branches and effects of spatial change of VEGF concentration. 
Although this model qualitatively reproduced the effects of VEGF concentration and the cell division of ECs, it at the same time produced unphysical phenomena such as a rapid increase in the density of ECs and the stop of bifurcation in weak proliferation.
It also did not incorporate branch reconnection, which is a particularly important element in two dimensions in determining the shape of the network.

In the present paper, we consider a modified model in which the temporal development of the vascular network is assumed to be determined by the mean EC density of the growing branches, and in which the reconnection of the vessels is taken into account.
We will make a number of assumptions to facilitate the analysis of the model.
Firstly, we assume that the vascular network extends in a two dimensional plane, like the retinal blood vessels and those in the in vitro experiments in Ref.~\cite{Arima-Nishiyama}. 
The ECs are initially supplied by an existing blood vessel, with a later contribution from cell division in the vascular network.
We assume that a branch elongates when the average density of ECs per length around the tip is greater than $\rho_e$, and bifurcates when it reaches the value $\rho_b$ ($\rho_b>\rho_e$).  
Let the length and cross section of the first neogenetic blood vessel be $L_0$ and $s_0$ respectively, and it bifurcates at $t=t_0:=0$. 
Since a blood vessel bifurcates into only two branches, we put the cross sections of the new branches $s_1^{(1)}$, and $s_1^{(2)}$.
According to Murray's law\cite{Murray14},   
\[
s_0^\mu=(s_1^{(1)})^\mu+(s_1^{(2)})^\mu \quad  ( 1. 4 \lesssim \mu \lesssim 1.6).
\]
We denote $\theta_1^{(i)}$ ($i=1,2$) by the angle of the direction of $i$th new branch to that of the original branch.  
In the previous model, we set $\theta_1^{(1)}=-\theta_1^{(2)}=\frac{\pi}{6}$, but we will consider different possibilities. 
The EC density of a branch is assumed not to change after bifurcation, that is, it remains to $\rho_b$.
We also assume that the average EC density in each growing branch is equal at the same time.
Due to this assumption and that a blood vessel bifurcates when the EC density reaches $\rho_b$, bifurcation takes place simultaneously in branches of the same generation. 
Similar to the previous model, the length of the $i$th branch $L_1^{(i)}(t)$ ($i=1,2$) satisfies 
\begin{equation}\label{1st_branch_eq}
\frac{d L_1^{(i)}}{d t}(t)=\lambda_1^{(i)}F(\rho_1(t))\quad (t>0),
\end{equation}   
where the parameter $\lambda_1^{(i)}$ denotes the effect of the environment around the $i$th branch, and $F$ is a non-decreasing function of the average EC density.
Let  $N_1(t)$ be the total number of ECs in the two growing branches.
Clearly,
\begin{equation}\label{1st_row_L_eq}
N_1(t)=\rho_1(t)\left(s_1^{(1)}L_1^{(1)}(t)+s_1^{(2)}L_1^{(2)}(t) \right),
\end{equation}
and
\begin{equation}\label{L1_ratio}
\frac{L_1^{(1)}(t)}{L_1^{(2)}(t)}=\frac{\lambda_1^{(1)}}{\lambda_1^{(2)}} \quad ({}^\forall t>0).
\end{equation}
The total volume of the growing vessels, $V_1(t)$, is given by
\begin{equation}\label{1st_volume}
V_1(t)=s_1^{(1)}L_1^{(1)}(t)+s_1^{(2)}L_1^{(2)}(t).
\end{equation} 
From Eqs.~\eqref{1st_branch_eq} and \eqref{1st_volume}, we have 
\begin{align*}
\frac{d V_1}{d t}(t)&=\Lambda_1 F(\rho_1(t)),\\
\Lambda_1&:=s_1^{(1)}\lambda_1^{(1)}+s_1^{(2)}\lambda_1^{(2)}.
\end{align*}
Using Eqs.~\eqref{1st_row_L_eq} and \eqref{1st_volume}, we obtain
\begin{equation}\label{V1_evolution}
\frac{d V_1}{d t}(t)=\Lambda_1 F\left(\frac{N_1(t)}{V_1(t)}\right),
\end{equation}
or, equivalently,
\begin{equation}\label{1st_rho_N_eq}
\frac{d}{d t}\left(\frac{N_1(t)}{\rho_1(t)} \right)=\Lambda_1 F \left( \rho_1(t) \right).
\end{equation}
Note that 
\[
N_1(+0)=V_1(+0)=0, \quad \rho_1(+0)=\lim_{t \to +0}\frac{N_1(t)}{V_1(t)}.
\]
Let us define
\begin{equation}
\nu_1:=\lim_{t \to +0}\frac{d N(t)}{d t}.
\end{equation}
We have an intrinsic relation to determine $\rho_1(+0)$ as
\begin{equation}\label{rho_1_0}
\frac{\nu_1}{\Lambda_1}= \rho_1(+0) F(\rho_1(+0)).
\end{equation}
Making a plausible assumption 
$\rho_1(+0)<\rho_b$, 
it holds that
\begin{equation}\label{inequality_rho_b}
\frac{\nu_1}{\Lambda_1} < \rho_b F(\rho_b),
\end{equation}
because $F(\rho_b)>0$ and $F$ is a non-decreasing function.

Suppose that bifurcation takes place at $t=t_1$, which implies $\rho_1(t_1)=\rho_b$ and we can determine the critical time $t_1$ from \eqref{1st_rho_N_eq} for a given $N_1(t)$.
The lengths of the branches are given as
\[
L_1^{(i)}:=L_1^{(i)}(t_1)=\frac{\lambda_1^{(i)} N_1(t_1)}{\rho_b\left(s_1^{(1)}\lambda_1^{(1)}+s_1^{(2)}\lambda_1^{(2)}\right) }=\frac{\lambda_1^{(i)} N_1(t_1)}{\rho_b\Lambda_1}   \quad (i=1,2).
\]
Then, each branch bifurcates into two new branches, and  for $t>t_1$ there are $2^2=4$ branches to grow.
Let the length and cross section of $i$th new branch be $L_2^{(i)}(t)$ and $s_2^{(i)}$ respectively ($i=1,2,3,4$).
The time evolution equation is 
\[
\frac{d L_2^{(i)}}{d t}=\lambda_2^{(i)}F(\rho_2(t)) \quad (i=1,2,3,4),
\]
where $\lambda_2^{(i)}$ is an environment parameter, $\rho_2(t)$ is the average EC density of the 4 growing branches, and, for $t>t_2$, 
\[
\rho_2(t)=\frac{N_2(t)}{V_2(t)}, \quad
V_2(t)=\sum_{i=1}^4 s_2^{(i)}L_2^{(i)}(t),
\]
with $N_2(t_1)=0$, $L_2^{(i)}(t_1)=0$ ($i=1,2,3,4$). 
Determining the critical time $t_2$ from $\rho_2(t_2)=\rho_b$, we obtain the branch length $L_2^{(i)}:=L_2^{(i)}(t_2)$ as
\[
L_2^{(i)}=\frac{\lambda_2^{(i)} N_2(t_2)}{\rho_b \Lambda_2 }\quad (i=1,2,3,4),
\]
where
\[
\Lambda_2:=\sum_{j=1}^4 s_2^{(j)}\lambda_2^{(j)}.
\]
In general, $k$th bifurcation takes place at $t=t_{k-1}$ and $2^k$ new branches arise and grow.
The equations for the growth of the $i$th branch ($i=1,2,...,2^k$) for $t>t_{k-1}$ are given in the same way as
\begin{subequations}
\begin{align}
\frac{d L_k^{(i)}}{d t}&=\lambda_k^{(i)}F(\rho_k(t)) \quad (i=1,2,\ldots,2^k), \label{sec2_difeq_a}\\
\rho_k(t)&=\frac{N_k(t)}{V_k(t)}, \label{sec2_difeq_b}\\
V_k(t)&=\sum_{i=1}^{2^k} s_k^{(i)}L_k^{(i)}(t), \label{sec2_difeq_c}\\
\frac{d V_k}{d t}(t)&=\Lambda_k F(\rho_k(t)),\label{sec2_difeq_d}
\end{align} 
\end{subequations}
where $L_k^{(i)}(t)$, $N_k(t)$, $s_k^{(i)}$ are the length of $i$th growing branch, total number of ECs in the growing branches, the cross section of $i$th branch respectively, and satisfy
\begin{equation}
{}^\forall i \; L_k^{(i)}(t_{k-1})=0,\quad N_k(t_{k-1})=0.
\end{equation}
Then, at $t=t_k$, $\rho_k(t_k)=\rho_b$, and $k+1$th bifurcation occurs.
The final length $L_k^{(i)}:=L_k^{(i)}(t_k)$ is given as
\begin{equation}\label{sec2_Lki}
L_k^{(i)}=\frac{\lambda_k^{(i)} V_k(t_k)}{\Lambda_k}=\frac{\lambda_k^{(i)} N_k(t_k)}{\rho_b \Lambda_k}\quad (i=1,2,\ldots,2^k),
\end{equation}
where
\begin{equation}
\Lambda_k:=\sum_{j=1}^{2^k} s_k^{(j)}\lambda_k^{(j)}.
\end{equation}
By solving the simultaneous equations \eqref{sec2_difeq_a} -- \eqref{sec2_difeq_a}, we obtain a set of lengths of branches in a vascular network.
\begin{figure}[htbp]
\begin{center}
\includegraphics[scale=0.7]{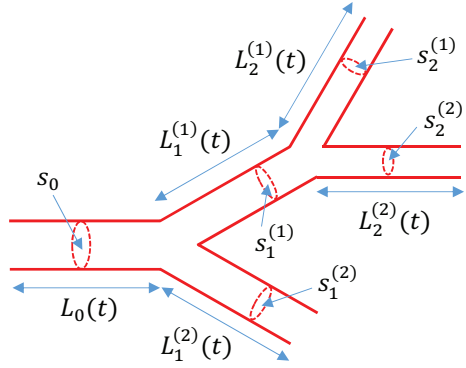}
\caption{Length of the branch $L_k^{(i)}$ and cross section $s_k^{(i)}$ in sprouting neogenetic blood vessel.}
\label{fig:euler}
\end{center}
\end{figure}
%

In the discrete dynamical equation model for angiogenesis in Ref~\cite{MYMKT}, the elongation and bifurcation are
determined by an algorithm based on the number of ECs at the tip, and the angle between the two branches bifurcated at the same point is set to be $\frac{\pi}{3}$  
from experimental observations.
In Ref.~\cite{MMYKT}, as its continuous approximation, the elongation and bifurcation is assumed to be determined by the number of the ECs, but density of ECs is assumed to be always uniform in the whole vascular network.
The latter assumption leads to unrealistic situations, such as an unlimited increase in the density of ECs.
Hence, in the present model, we modified so that the density of ECs has an upper bound $\rho_b$.
The functions $F(\rho)$ and $N_k(t)$ ($k=1,2,...$) may be estimated from experimental results, but we do not have enough data.
A plausible assumption is that 
\begin{equation}
F(\rho)=\left\{
\begin{array}{cl}
\rho-\rho_e&\ (\rho \ge \rho_e)\\
0&\ (\rho <\rho_e)
\end{array}
\right. ,\label{sec2_F_form}
\end{equation}
where $\rho_e$ is the threshold density of elongation and a positive constant.
Since supply rate of ECs exponentially grows at first, but will stop after sufficient number of  ECs are supplied,  
the total number of ECs, $N(t)$, may be assumed to satisfy the following Logistic equation:
\begin{equation}\label{Logistic}
\frac{d}{dt}N(t)=\left(\epsilon - bN(t)\right)N(t) \quad (N(t_0)=N_{ini}),
\end{equation}
where 
\[
N(t):=\sum_{k=0}^\infty N_k(t)\quad \left(\mbox{$N_0(t)=N_{ini}$, ${}^\forall k,\, N_k(t):=0$ ($t<t_{k-1}$)}\right),
\]
 $\epsilon$ and $b$ are positive constants and $N_\infty:=\frac{\epsilon}{b}>N_{ini}$ is the final total number of ECs in the network.
Note that Eq.~\eqref{Logistic} has a unique solution 
\begin{equation}\label{Logistic_sol}
N(t)=\frac{N_\infty}{1+\left(\frac{N_\infty}{N_{ini}}-1 \right)\e^{-\epsilon t}}.
\end{equation}

\begin{figure}[htbp]
\begin{center}
\includegraphics[scale=0.7]{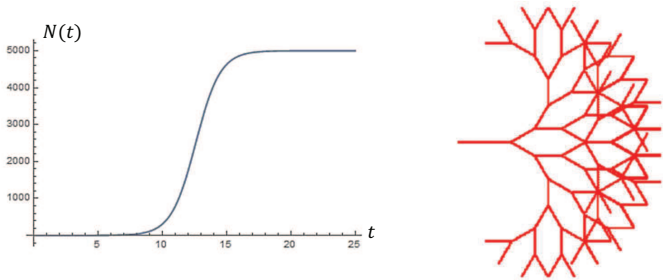}
\caption{Bifurcation of the present model when $N(t)$ is given by \eqref{Logistic} or, equivalently,\eqref{Logistic_sol}. 
The parameters are $\rho_e=18,\ \rho_b=20,\ \epsilon=1.05,\, b=1.05\times 10^{-4},\, N_{ini}=10^{-3}$ and we assume the Murray's law ($\mu=1.5$).
Bifurcation angle $\theta$ is $\frac{\pi}{6}$ at every bifurcation.}
\label{Mada_fig_2}
\end{center}
\end{figure}

Figure~\ref{Mada_fig_2} is an example of the network pattern obtained in the present model.
The functions $F(\rho)$ and $N(t)$ are given by Eqs.~\eqref{sec2_F_form} and \eqref{Logistic}.
The angle between the direction of a bifurcated branch and the original branch, to which hereafter we will refer as the \textit{bifurcation angle}, is set to be $\frac{\pi}{6}$.
There are many branches crossing each other.
We investigate the effects of reconnection in the next section. 
 
%

The functions $F(\rho)$ and $N_k(t)$ are non-decreasing functions, but they are not necessarily restricted to those in Eqs.~\eqref{sec2_F_form} and \eqref{Logistic}.
Let us consider several cases where we can analytically estimate time evolution of the growing branches.
\begin{description}
\item[(A)] $F(\rho)=f_0\, (const.)$.

From \eqref{V1_evolution} and \eqref{1st_rho_N_eq}, we have
\[
V_1(t)=\Lambda_1 f_0 t,\quad
\rho_1(t)=\frac{N_1(t)}{\Lambda_1 f_0 t}.
\]
Since $\rho_1(+0)=\frac{1}{\Lambda_1 f_0}\frac{d}{dt}N_1(+0)=\frac{\nu_1}{\Lambda_1 f_0}$, it holds that 
\[
\frac{\nu_1}{\Lambda_1 f_0}<\rho_b.
\]
Let bifurcation take palce at $t=t_1$, which implies $\rho_1(t_1)=\rho_b$.
Hence, for a given $N_1(t)$, we can determine the critical time $t_1$ and obtain the lengths of the branches as
\[
L_1^{(i)}=\lambda_1^{(i)}f_0 t_1 \quad (i=1,2).
\]
In general,  putting $\Delta t_k:=t_k-t_{k-1}$, 
\begin{equation}\label{L_caseA}
L_k^{(i)}=\lambda_k^{(i)}f_0\Delta t_k \quad (i=1,2,...,2^k).
\end{equation}

To determine the critical time $t_k$, we have to give the number of ECs supplied to the growing branches.
Let us consider two typical examples.
\begin{itemize}
\item If $N_k(t)$ ($k=1,2,...$) increases exponentially as 
\[
N_k(t)=\frac{\nu_k}{a_k}\left(\e^{a_k (t-t_{k-1})}-1 \right),
\]
Since $N_k(t_k)=\rho_bV_k(t_k)=\rho_b\sum_is_k^{(i)}L_k^{(i)}(t_k)$, $\Delta t_k$ is determined from the transcendental equation
\[
\frac{\nu_k}{a_k}\left(\e^{a_k\Delta t_k}-1 \right)=\rho_b \Lambda_k f_0 \Delta t_k,
\]
and, from \eqref{L_caseA}, we obtain the length of $L_k^{(i)}$.
\item If it is given by quadratic form with a positive constant $a_k$:
\[
N_k(t)=\nu_k(t-t_{k-1})\left(1+a_k(t-t_{k-1})\right),
\]
we have
\[
\frac{N_k(t_k)}{V_k(t_k)}=\frac{\nu_k\Delta t_k(1+a_k \Delta t_k)}{f_0\Lambda_k \Delta t_k}=\rho_b
\]
and
\[
\Delta t_k=\frac{\rho_b f_0 \Lambda_k-\nu_1}{a_k \nu_k}.
\]
Hence we find explicit formula for $L_k^{(i)}$ as
\begin{equation}\label{sec2_A_2}
L_k^{(i)}=\frac{\lambda_k^{(i)}f_0 \left(\rho_b f_0 \Lambda_k-\nu_k\right)}{a_k \nu_k}\quad (i=1,2,...,2^k).
\end{equation}
\end{itemize}

\item[(B)] $F(\rho)=\rho^\mu$ ($\mu>0$).

Equation for $V_k(t)$ is written as
\[
\frac{d V_k}{d t}=\Lambda_k \left( \frac{N_k(t)}{V_k(t)}\right)^\mu.
\]
Thus, we have an integral
\[
\Lambda_k \int_{t_{k-1}}^{t} N_k^\mu (t')\,dt'=\int_0^{V_k(t)}V_k^\mu \,dV_k=\frac{1}{1+\mu}V_k(t)^{\mu+1}. 
\]
Hence, for a given $N_k(t)$, we obtain 
\[
V_k(t)=\left[ (1+\mu)\Lambda_k \int_{t_{k-1}}^t (N_k(t'))^\mu\, dt'  \right]^{\frac{1}{1+\mu}},
\]
and $t_k$ is determined by  $N_k(t_k)=\rho_b V_k(t_k)$.
\begin{itemize}
\item If $N_k(t)$ increases exponentially, 
\[
N_k(t)=\frac{\nu_k}{a_k}\left( \e^{a_k(t-t_{k-1})}-1\right)\quad (t>t_{k-1}),
\]
then 
\[
\frac{\nu_k}{a_k}\left( \e^{a_k \Delta t_k}-1\right)=\rho_b\left[ (1+\mu)\Lambda_k \left(\frac{\nu_k}{a_k}\right)^\mu \int_0^{\Delta t_k} (\e^{a_k t'}-1)^\mu\, dt'  \right]^{\frac{1}{1+\mu}},
\]
or
\[
\int_0^{\Delta t_k}\left(\e^{a_kt'}-1 \right)^\mu\, dt' = \frac{\nu_k}{(1+\mu)\rho_b^{1+\mu}a_k\Lambda_k}\left(\e^{a_k\Delta t_k}-1 \right)^{1+\mu}.
\]
Since
\[
\int_0^{\Delta t_k}\left(\e^{a_kt'}-1 \right)^\mu\, dt' 
=\frac{(\e^{a_k\Delta t_k}-1)^{\mu+1}}{(1+\mu)a_k}{}_2F_1
\left(1,1+\mu,2+\mu; 1-\e^{a_k\Delta t_k}   \right),
\]
we find a transcendental equation to determine $\Delta t_k$ as
\[
{}_2F_1
\left(1,1+\mu,2+\mu; 1-\e^{a_k\Delta t_k}   \right)=\frac{\nu_k}{\rho_b^{1+\mu} \Lambda_k}.
\]
Here ${}_2F_1(\alpha,\beta,\gamma;x)$ is the hypergeometric function:
\[
{}_2F_1(\alpha,\beta,\gamma;x)=\sum_{n=0}^\infty \frac{(\alpha)_n(\beta)_n}{(\gamma)_nn!}x^n.
\]
For example, in case $\mu=2$, 
\begin{align*}
&{}_2F_1
\left(1,1+\mu,2+\mu; 1-\e^{a_k\Delta t_k}   \right)\big|_{\mu=2}\\
&=\frac{3\left\{a_k\Delta t_k-(\e^{a_k\Delta t_k}-1)+ \frac{(\e^{2a_k\Delta t_k}-1)^2}{2}\right\}}{(\e^{a_k\Delta t_k} -1)^3}.
\end{align*}
\item If $N_k(t)$ increases quadratically;
\[
N_k(t)=\nu_k(t-t_{k-1})\left(1+a_k(t-t_{k-1})\right),
\]
we have
\begin{align*}
\nu_k\Delta t_k \left(1+a_k\Delta t_k\right)&=\rho_b
\left[(1+\mu)\Lambda_k \nu_k^\mu  \int_0^{\Delta t_k} \left(t'(1+a_kt')\right)^\mu\,dt'    
\right]^{\frac{1}{1+\mu}}\\
&=\rho_b\left[\Lambda_k \nu_k^\mu (\Delta t_k)^{\mu+1}{}_2F_1(-\mu,1+\mu,2+\mu;\,-a_k\Delta t_k)
\right]^{\frac{1}{1+\mu}},
\end{align*}
or
\[
(1+a_k\Delta t_k)^{1+\mu}=\frac{\rho_b^{1+\mu}\Lambda_k}{\nu_k}{}_2F_1(-\mu,1+\mu,2+\mu;\,-a_k\Delta t_k).
\]
For example, in case $\mu=1$,
\[
{}_2F_1(-\mu,1+\mu,2+\mu;\,-a_k\Delta t_k)\big|_{\mu=1}
=1+\frac{2a_k\Delta t_k}{3}.
\]
Assuming $\rho_b^2\Lambda_k>\nu_k$ so that the condition $\Delta t_k>0$ is satisfied, we have
\[
\Delta t_k=\frac{-3\nu_k+\rho_b^2\Lambda_k+\sqrt{\rho_b^4\Lambda_k^2+3\nu_k\rho_b^2\Lambda_k}}{3\nu_k a_k},
\]
and, from \eqref{sec2_Lki}, 
\[
L_k^{(i)}=\lambda_k^{(i)}\frac{2\Lambda_k^2\rho_b^3+(2\Lambda_k\rho_b^2-3\nu_k)\sqrt{\Lambda_k^2\rho_b^2+3\Lambda_k\nu_k}}{9 a_k \nu_k\Lambda_k}.
\]
\end{itemize}
\item[(C)] In case EC supply rate is constant, that is, $N_k(t)=\nu_k (t-t_{k-1})$.

Since $V_k(t)=\frac{N_k(t)}{\rho_k(t)}$, we have
\[
\frac{d V_k}{d t}=\frac{d}{dt}\left(\frac{N_k(t)}{\rho_k(t)} \right)=\frac{\nu_k}{\rho_k(t)}
-\frac{\nu_k (t-t_{k-1})}{(\rho_k(t))^2}\frac{ d \rho_k}{d t}(t).
\]
Hence we have the differential equation
\begin{equation}\label{linear_case}
(t-t_{k-1})\frac{d \rho_k}{d t}=\rho_k-\frac{\Lambda_k}{\nu_k}\rho_k^2F(\rho_k),
\end{equation}
which is integrable as
\[
\int_{\rho_k(t_{k-1})}^{\rho_k(t)}\frac{d \rho}{\rho-\frac{\Lambda_k}{\nu_k}\rho^2F(\rho)}=\int_{t_{k-1}}^t \frac{dt'}{t'-t_{k-1}}.
\]
Note that $\rho_k(t_{k-1})$ is given by
\begin{equation}\label{rho_0_value}
\rho_k(t_{k-1})-\frac{\Lambda_k}{\nu_k}\rho_k(t_{k-1})^2F(\rho_k(t_{k-1}))=0.
\end{equation}
By taking limit,
\[
\frac{d \rho_k}{d t}(t_{k-1})=\lim_{t\to t_{k-1}}\frac{\rho_k(t)-\frac{\Lambda_k}{\nu_k}\rho_k(t)^2F(\rho_k(t))}{t-t_{k-1}},
\]
we also find that, if $\frac{d \rho_k}{dt}(t_{k-1})\ne 0$,  
\begin{equation}\label{dif_rho_condition}
2\rho_k(t_{k-1})F(\rho_k(t_{k-1}))+\rho_k(t_{k-1})^2F'(\rho_k(t_{k-1}))=0.
\end{equation}
Since $F(\rho)$ is a non-decreasing function, we have $\rho_k(t_{k-1})= 0$.
In the present case, to satisfy this condition, we do not assume that $L_k^{(i)}(t_{k-1})=0$, but that $V_k(t_{k-1}):=V_0>0$, that is, we require the existence of small buds of the new branches at $t=t_{k-1}+0$.
\begin{itemize}
\item The case in which $F$ is given as 
\[
F(\rho)=\left\{
\begin{array}{cl}
\rho-\rho_e&\quad (\rho \ge \rho_e)\\
0 &\quad (\rho < \rho_e)
\end{array}
\right. ,
\] 
where $\rho_e$ is a positive constant.

Since $\rho(t_{k-1})=0$, no elongation takes place until the time $t_{k-1}'$ at which the density $\rho_k$ reaches $\rho_e$:
\[
\frac{\nu_k(t_{k-1}'-t_{k-1})}{V_0}=\rho_e.
\]

Using the formula:
\[
\int \frac{dx}{x(1+\alpha x)(1-\beta x)}=\log \left|x(1+\alpha x)^{-\frac{\alpha}{\alpha+\beta}}(1-\beta x)^{-\frac{\beta}{\alpha+\beta}}\right|, 
\]
we obtain
\[
\int\frac{d \rho}{\rho-\frac{\Lambda_k}{\nu_k}\rho^2F(\rho)} 
=\log \left| \rho(1+\alpha \rho)^{-\frac{\alpha}{\alpha+\beta}}
(1-\beta\rho )^{-\frac{\beta}{\alpha+\beta}}  \right|,
\]
where positive values $\alpha$, $\beta$ are determined by
\[
\alpha-\beta=\frac{\Lambda_k\rho_e}{\nu_k},\quad \alpha\beta=\frac{\Lambda_k}{\nu_k}.
\]
Hence we find that
\[
\frac{\rho_b(1+\alpha \rho_b)^{-\frac{\alpha}{\alpha+\beta}}
(1-\beta\rho_b )^{-\frac{\beta}{\alpha+\beta}} }{\rho_e(1+\alpha \rho_e)^{-\frac{\alpha}{\alpha+\beta}}
(1-\beta\rho_e )^{-\frac{\beta}{\alpha+\beta}} } 
=\frac{t_k-t_{k-1}}{t_{k-1}'-t_{k-1}}, 
\]
and
\begin{equation}\label{sec2_T_cvalue}
\Delta t_k
=\frac{V_0}{\nu_k}\frac{\rho_b(1+\alpha \rho_b)^{-\frac{\alpha}{\alpha+\beta}}
(1-\beta\rho_b )^{-\frac{\beta}{\alpha+\beta}} }{(1+\alpha \rho_e)^{-\frac{\alpha}{\alpha+\beta}}
(1-\beta\rho_e )^{-\frac{\beta}{\alpha+\beta}} } .
\end{equation}
With this expression, the branch length is given by
\begin{equation}\label{sec2_L_cvalue}
L_k^{(i)}=\frac{\lambda_k^{(i)}\nu_k \Delta  t_k}{\Lambda_k \rho_b}.
\end{equation}
\item A more general case in which $F$ is expressed as a rational function as
\[
F(\rho)=\left\{
\begin{array}{cl}
\frac{\rho-\rho_e}{\rho+c} &\quad (\rho \ge \rho_e)\\
0 &\quad (\rho < \rho_e)
\end{array}
\right. ,
\] 
where $\rho_e$ and $c$ are positive constants.
Without loss of generality, we take $c=1$.

As in the previous case, we consider the time $t_{k-1}'$ which satisfies
\[
\frac{\nu_k(t_{k-1}'-t_{k-1})}{V_0}=\rho_e.
\]
Using the formula:
\[
\int \frac{(1+x) dx}{x(1+\alpha x)(1-\beta x)}=\log \left|x(1+\alpha x)^{\frac{1-\alpha}{\alpha+\beta}}(1-\beta x)^{-\frac{1+\beta}{\alpha+\beta}}\right|, 
\]
we can estimate $\Delta t_k$ in the same way, and obtain
\[
\Delta t_k
=\frac{\nu_k}{\rho_e V_0}\frac{\rho_b(1+\alpha' \rho_b)^{\frac{1-\alpha'}{\alpha'+\beta'}}
(1-\beta'\rho_b )^{-\frac{1+\beta'}{\alpha'+\beta'}} }{\rho_e(1+\alpha' \rho_e)^{\frac{1-\alpha'}{\alpha'+\beta'}}
(1-\beta'\rho_e )^{-\frac{1+\beta'}{\alpha'+\beta'}} },
\]
where
\[
\alpha'-\beta'=1+\frac{\Lambda_k\rho_e}{\nu_k},\quad \alpha'\beta'=\frac{\Lambda_k}{\nu_k}.
\]
The branch length $L_k^{(i)}$ is obtained similarly.

\end{itemize}
\end{description}

To close this section, let us consider some special cases in which the branching ratio becomes constant.
Suppose that all the cross sections of new branches arose from the $k$th bifurcation ($k=1,2,...$) are the same, that is, ${}^\forall i,\, s_k^{(i)}=s_k$, we
find from the Murray's law that
\[
s_k=(2^{-\frac{1}{\mu}})^ks_0.
\]
We also assume that all the environmental parameters are common $\lambda_k^{(i)}=\lambda$, and that the supply rate of ECs ($\nu_k$) just after the bifurcation is proportional to $\Lambda_k$, $\Lambda_k=\gamma \nu_k$, which essentially means that $\nu_k$ is proportional to the sum of the cross sections.

In case (A) with $N_k(t)=\nu_k(t-t_{k-1})(1+a_k(t-t_{k-1}))$, the coefficient $a_k$ denotes the increase of supply rate due to proliferation of ECs in new blood vessel branches.
Hence it will increase exponentially;
\[
a_k=a_0R^k\qquad (k=1,2,...).
\]
Then, from \eqref{sec2_A_2}, 
we find that 
\begin{equation}\label{sec2_proportional1}
L_k^{(i)}=R^{-k}L_0^*,
\end{equation} 
where $L_0^*:=\frac{\lambda f_0 \left(\rho_b f_0 \gamma-1\right)}{a_0}$.

While, in case (C) with $F=\rho$ ($\rho_e=0$),  the $i$th branch of $k$th bifurcation 
is given as 
\[
L_k^{(i)}=\frac{\lambda_k^{(i)}V_0}{\Lambda_k \sqrt{1-\frac{\Lambda_k}{\nu_k}\rho_b^2}}.
\]
Under the above assumptions, we have
\[
L_k^{(i)}=\frac{V_0}{s_0\sqrt{1-\gamma\rho_b^2}}2^{(\frac{1}{\mu}-1)k},
\]
which leads to
\begin{equation}\label{sec2_proportional2}
L_k^{(i)}=2^{(\frac{1}{\mu}-1)k}L_0^*\qquad (k=0,1,2,...),
\end{equation}
where $L_0^*:=\frac{V_0}{s_0\sqrt{1-\gamma\rho_b^2}}$.
In both cases, the vascular networks show self-similar structures and fractal in appearance.

In general, for an appropriate choice of the function $F$ and other corresponding parameters, we have special cases where networks exhibit self-similar behaviour.
If the ratio of the cross sections of the two generated branches and that of the branch lengths are always the same at bifurcation, the network has self-similarity.
Although this assumption is somewhat artificial, it would lead to a fractal-like pattern which is helpful in discussing fractal nature of vascular networks.  
We shall examine the fractal-like structure of these patterns in section 4.
So far we have not discussed the reconnection of blood vessels and bifurcation angles, that is, the angles between the original blood vessel and newly generated branches at bifurcation.
We investigate their effects on the networks in the following section.

%
%
%
\section{Vascular network with reconnection}
\label{sec:R-network}
%
%

\begin{figure}[htbp]
\begin{center}
\includegraphics[scale=0.7]{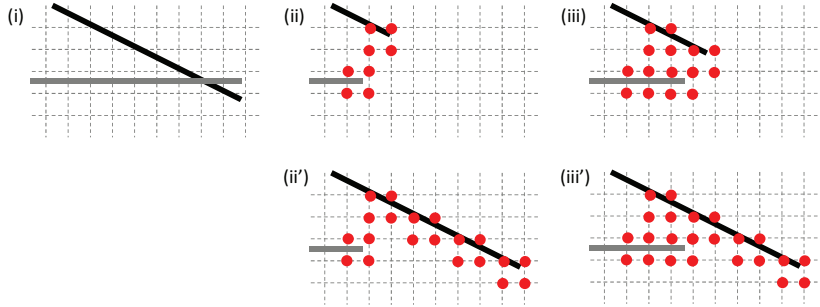}
\caption{Introducing an effect of reconnection into the numerical simulation. 
(i) In case no reconnection. 
The flow (ii) $\rightarrow$ (iii) shows the case where the black line segment and the grey line segment approach to touch. 
The flow (ii') $\rightarrow$ (iii') shows the case where the black line segment already exists and the grey line segment approaches. 
The points in the figures show the vertices of squares belonging to the branches.}
\label{fig:coupling}
\end{center}
\end{figure}

\begin{figure}[hbtp]
\begin{center}
\includegraphics[scale=0.5]{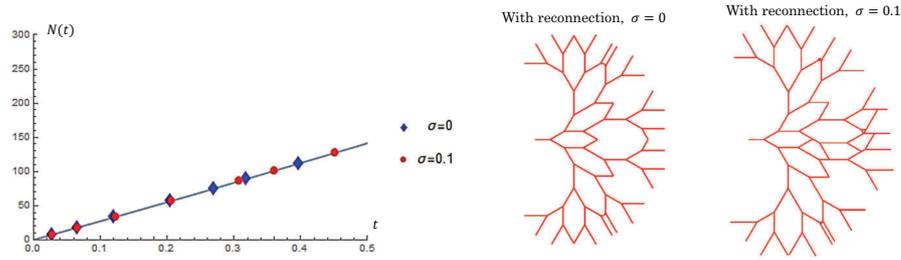}
\caption{Vascular networks with reconnection of blood vessels.
The left graph shows the total number of ECs $N(t)$ and the dots on the graph denote the points at which bifurcations take place. 
The parameters are set to $d=0.01$, $\rho_e=2, \rho_b=20$, and 
$N(t)=\frac{6030}{1+e^{-0.3(t - 5)}}-1100$.}
\label{fig:BVwithCE1}
\end{center}
\end{figure}

In this section, we consider effects of reconnection for the differential equation model given in the previous section. 
We adopt the function given in Eq.~\eqref{sec2_F_form} as  the function $F(\rho)$. 
As for $N(t)$, in order to adjust the position of the first branch, we use a parallel shift of the function  \eqref{Logistic_sol} (See the capsion of Fig.~\ref{fig:BVwithCE1}). 
Although the model is expressed in terms of non-autonomous simultaneous ordinary differential equations, it is fairly difficult to find the intersection places succinctly.
We express the branches as line segments and do not take the width of the blood vessels into account to determine the reconnection places. 
The bifurcation angle is considered as a given parameter.
Furthermore we divide the two dimensional plane, in which the vascular network is constructed, into a fine square lattice of width $d$.
When a line segment falls within a square, we consider that square to be occupied by a blood vessel corresponding to the line segment.
According to the differential equations, each branch length $L_k^{(i)}(t)$ is calculated numerically and the coordinate of the point of its tip is obtained for a given bifurcation angle.    
If we do not consider the reconnection of branches, we obtain a figure like (i) in  Fig.~\ref{fig:coupling}.
However, considering that the branches will reconnect when they cross, we have to treat two cases.
One is the case where the tips of the two branches approach and touch.
The other is the case where a tip of an elongating branch touches an existing branch. 
In the former case, we consider that the two branches reconnect when the corresponding two squares have the same vertex ((ii) to (iii) in  Fig.~\ref{fig:coupling}). 
In the latter case, an elongating branch reconnect to the existing branch when its tip square comes to have a vertex of a square belonging to the existing branch ((ii') to (iii') in  Fig.~\ref{fig:coupling}). 

\begin{figure}[hbtp]
\begin{center}
\includegraphics[scale=0.5]{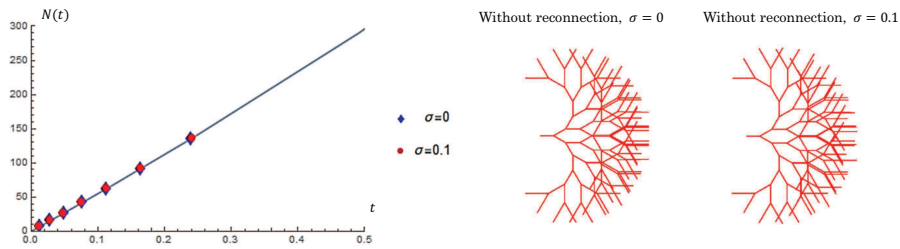}
\caption{Vascular networks without reconnection of blood vessels. The parameters are the same as those in Fig.~\ref{fig:BVwithCE1}.}
\label{fig:BVwithCE2}
\end{center}
\end{figure}

By means of the above procedure, we numerically obtain vascular network patterns as shown in Fig.~\ref{fig:BVwithCE1}.
We take the bifurcation angle $\theta=\frac{\pi}{6}$.
The parameter $\lambda_k^{(i)}$ is randomly determined so that its distribution follows a normal distribution with mean $1$ and variance $\sigma$.
The other parameters are written in the figure caption.
In comparison to Fig.~\ref{fig:BVwithCE1}, we show the corresponding vascular networks without reconnection in Fig.~\ref{fig:BVwithCE2}.

\begin{figure}[hbtp]
\begin{center}
\includegraphics[scale=0.6]{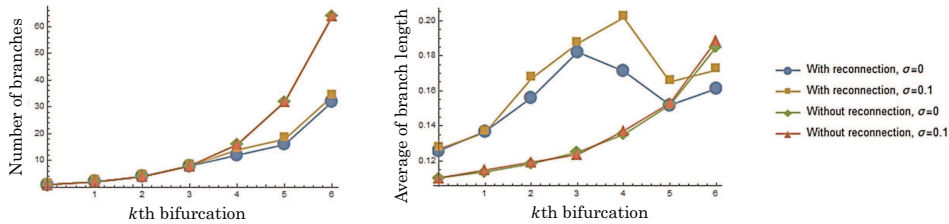}
\caption{The number of branches generated at each order of bifurcation and the average branch lengths of them in the Figs.~\ref{fig:BVwithCE1} and \ref{fig:BVwithCE2}. }
\label{fig:BVwithCE3}
\end{center}
\end{figure}

Figures~\ref{fig:BVwithCE3} show the total number of branches generated at each order of bifurcations and the average branch length of them. If no reconnection is made, the number of branches generated at $k$th bifurcation is $2^k$, but since reconnection eliminates branches, the increase in the number of branches is suppressed.
The average length of the branches increases without reconnection, but it is also reduced by reconnection. 

\begin{figure}[htbp]
\begin{center}
\includegraphics[scale=0.6]{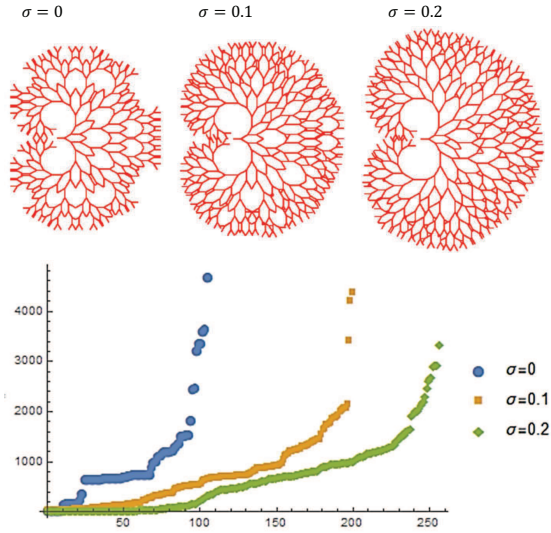}
\caption{Patterns and area distributions of islands that appear in the vascular networks when $\lambda_k^{(i)}$ follows a normal distribution of mean 1 and variance $\sigma\ (=0,\,0.1,\,0.2)$. 
The islands are numbered 1,2,3,... in order of smallest to largest area.
The vertical axis of the graph shows the area of the island, and the horizontal axis the number of the island.}
\label{fig:distribution}
\end{center}
\end{figure}

In order to see the characteristics of the vascular network thus created, we consider the distribution of the area of the islands in the vascular network as examined in Ref.~\cite{IJTYM}. 
An island here is a region in the network that is surrounded by blood vessels, but where no blood vessel is present. 
An area of an island is evaluated from the number of bits in the island by image analysis.
Figures~\ref{fig:distribution} show the area distribution of the islands for different values of variance $\sigma$.
The larger the variance value, the wider the distribution of the area of an island.

\begin{figure}[htbp]
\begin{center}
\includegraphics[scale=0.5]{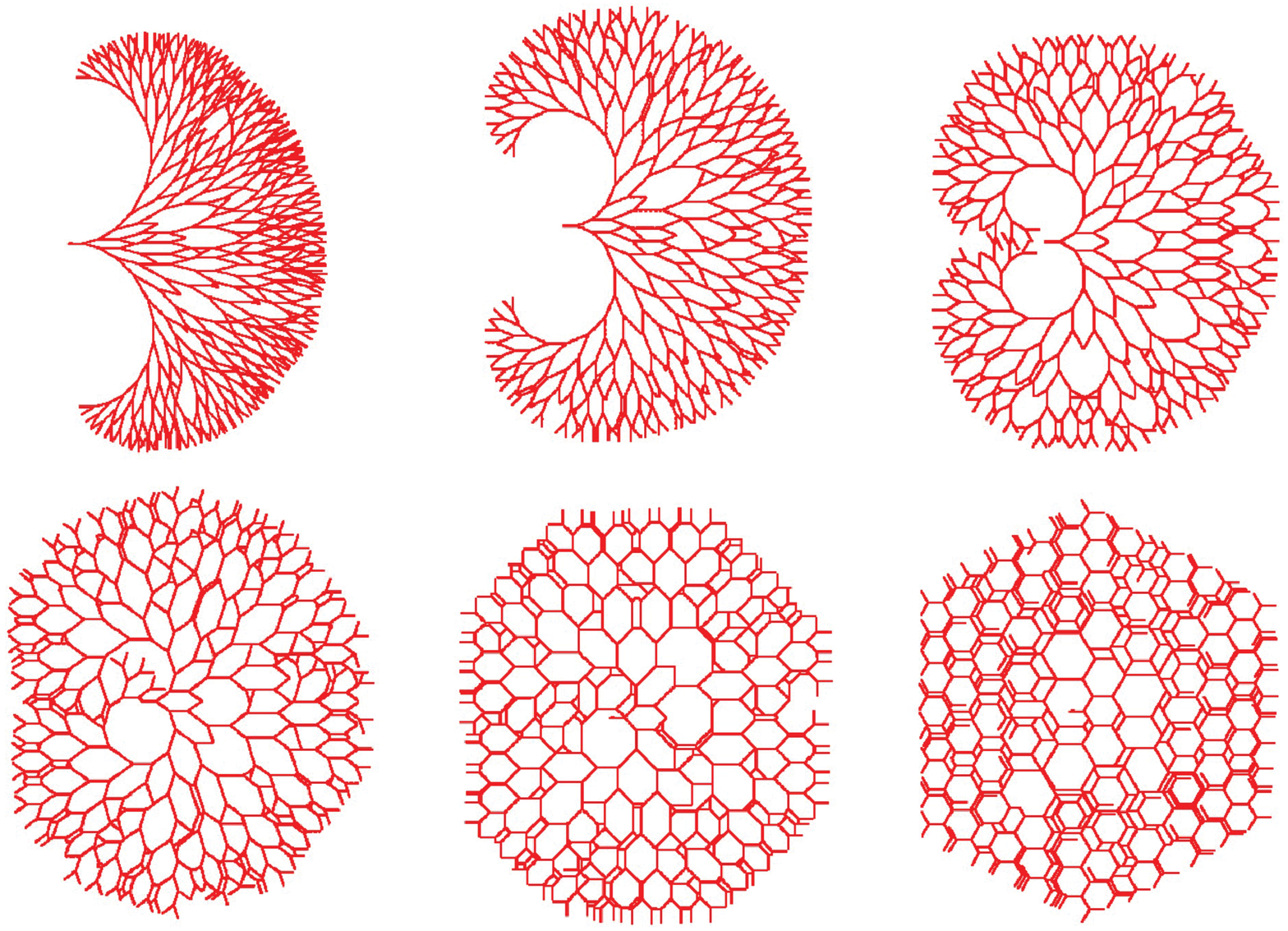}
\caption{The vascular networks with reconnection effects for various bifurcation angles. The bifurcation angles of the figures in the upper row are $\frac{\pi}{12}$, $\frac{\pi}{8}$, and $\frac{\pi}{6}$ from left to right, and $\frac{\pi}{5}$, $\frac{\pi}{4}$ 
and $\frac{\pi}{3}$ in the lower row. 
For $\theta=\frac{\pi}{3}$, the reason why there are areas where the hexagons appear to overlap is that some of branches are whiring in the sape of a hexagon. }
\label{fig:angle}
\end{center}
\end{figure}

\begin{figure}[htbp]
\begin{center}
\includegraphics[scale=0.45]{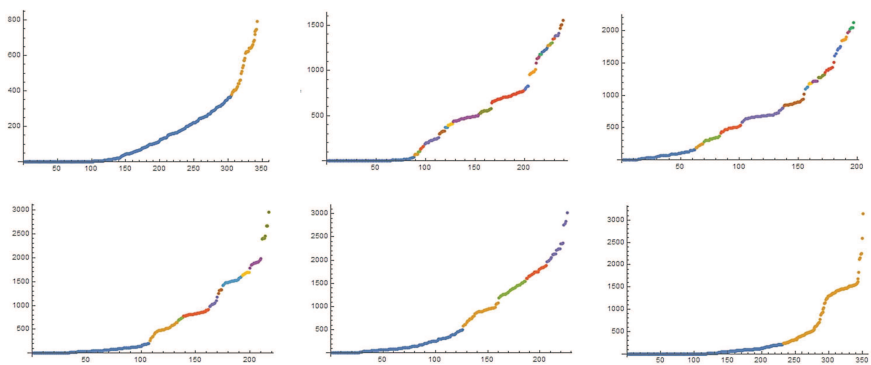}
\caption{Cluster analysis of size of islands for the patterns in Fig.~\ref{fig:angle}.
The vertical axis of the graphs is the size of the island, and the horizontal axis is the number of the island. (The islands are numbered 1,2,3,... in order of smallest to largest area as in Fig.~\ref{fig:distribution}.)
Outliers that are too large are ignored in the graph.}
\label{fig:cluster}
\end{center}
\end{figure}

\begin{figure}[htbp]
\begin{center}
\includegraphics[scale=0.45]{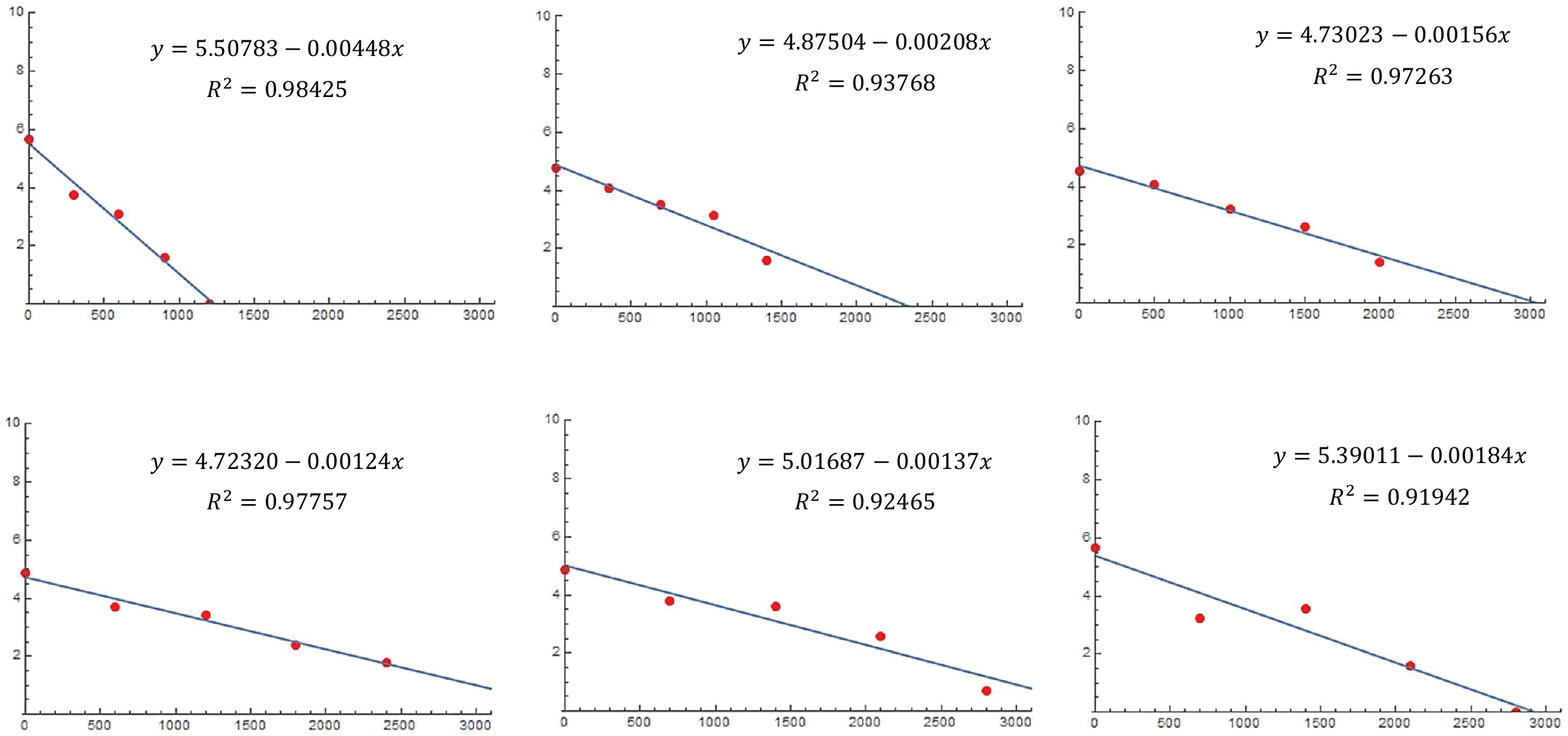}
\caption{Island size distribution for the patterns in Fig.~\ref{fig:angle}.
The vertical axis is the logarithmic value of frequency, and the horizontal axis is the size of the island. The straight lines show linear regressions.
These distributions, especially those for $\theta=\frac{\pi}{12},\, \frac{\pi}{6},$ and $\frac{\pi}{5}$, fit the regression lines well.}  
\label{fig:fitting}
\end{center}
\end{figure}

Next we consider pattern formation for different bifurcation angles.
Vascular networks  generated at $13$th bifurcation are shown in Fig.~\ref{fig:angle}. 
The corresponding distribution of area of the islands and the distribution map of the island size created based on this result are shown in the graphs of Fig.~\ref{fig:cluster} and Fig.~\ref{fig:fitting} respectively. 
Peculiar islands appear at the center and in the direction of angle $\pi$, but they are ignored in estimations in these graphs. 

The size distribution of the islands at angles $\frac{\pi}{6}$ and $\frac{\pi}{5}$ in Figs.~\ref{fig:cluster} and \ref{fig:fitting} shows properties similar to those in Ref.~\cite{IJTYM}, though the meaning of this fact is unclear.

\begin{figure}[htbp]
\begin{center}
\includegraphics[scale=0.4]{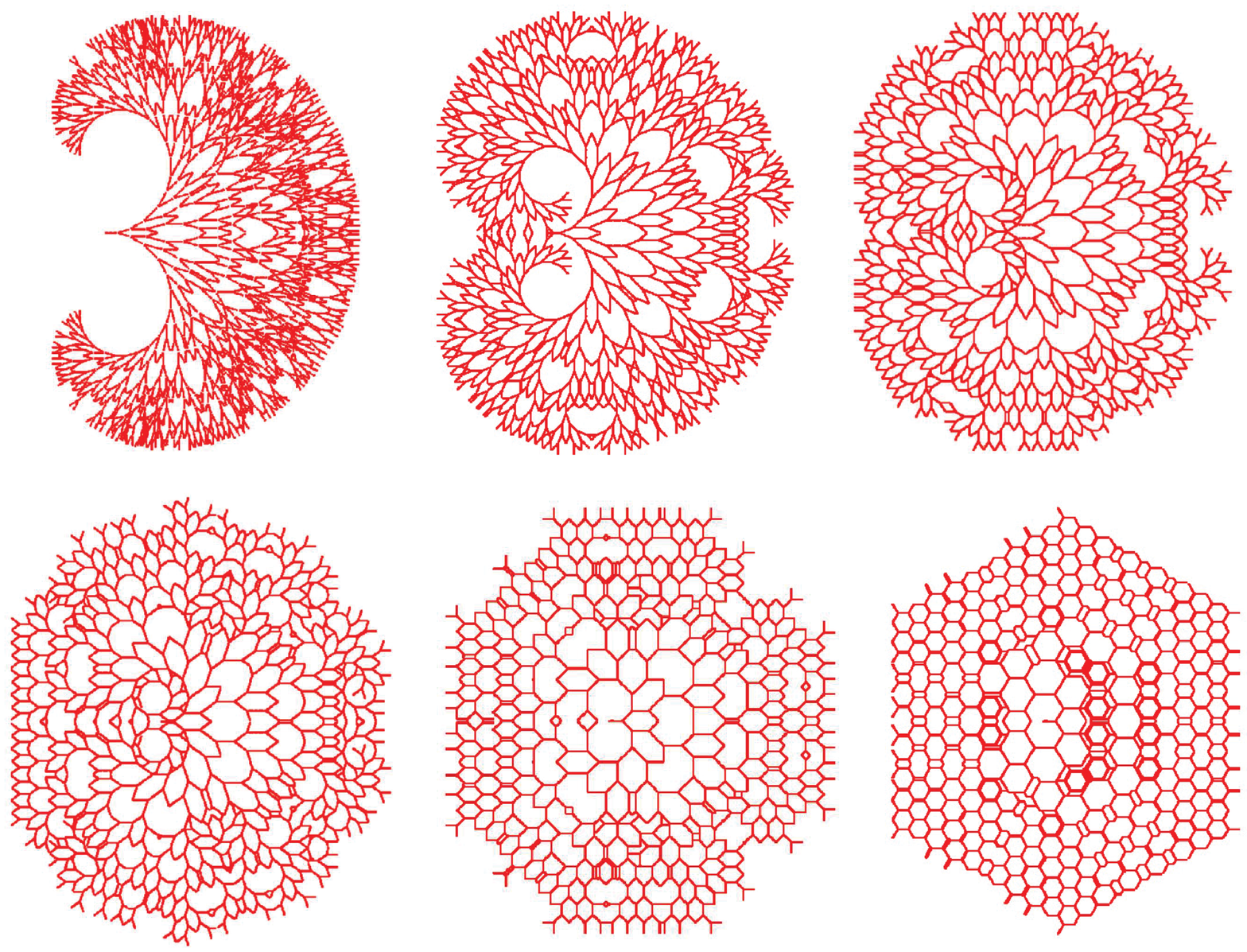}
\caption{Vascular network patterns used for the evaluation of area and length in Fig.~\ref{fig:AreaLength}. The variance $\sigma=0$ and up to $19$th-branch is calculated. The angles are $\frac{\pi}{12}$, $\frac{\pi}{8}$, and $\frac{\pi}{6}$ in the upper row from left to right, and $\frac{\pi}{5}$, $\frac{\pi}{4}$ 
and $\frac{\pi}{3}$ in the lower row.}
\label{fig:angle01v1}
\end{center}
\end{figure}

\begin{figure}[htbp]
\begin{center}
\includegraphics[scale=0.4]{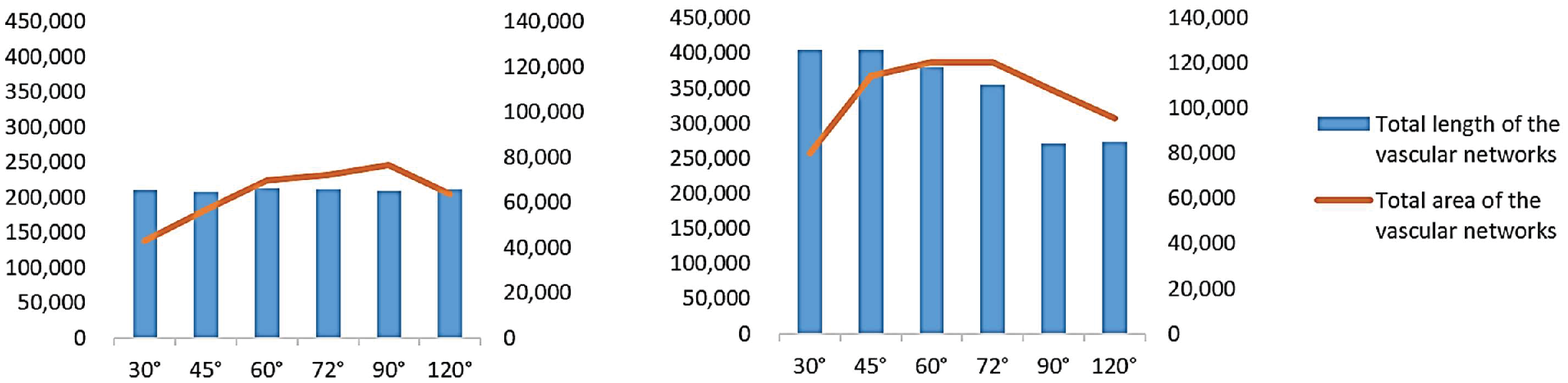}
\caption{Total area and total length of the vascular networks in Fig.~\ref{fig:angle} and Fig.~\ref{fig:angle01v1}.}
\label{fig:AreaLength}
\end{center}
\end{figure}

Finally we examine the dependence of the total length and total area of the vascular patterns on the bifurcation angle.
Figure~\ref{fig:AreaLength} show the total area and total length of the vascular networks in Fig.~\ref{fig:angle} and Fig.~\ref{fig:angle01v1}.  
The area is estimated by image analysis.
From the case of $19$th bifurcation, we see that the vascular network with bifurcation angle $\theta=\frac{\pi}{5}$ or $\frac{\pi}{6}$ covers wide area though the total length of branches is not so large.

%
\section{Properties of the model vascular network patterns}
\label{Chapter:fractal}
In this section, we consider the properties of network patterns in which the ratio of the length of the bifurcated branch to the original branch is a constant, $r\in \R_+$, as in the cases
\eqref{sec2_proportional1} or \eqref{sec2_proportional2}.
We also assume that the network develops in two dimensional plane like a vascular network in a retina, and that the angles between the direction of original branch to that of the two new bifurcated branches are $\pm \theta$ $\ (0<\theta <\frac{\pi}{2})$.
We call $\theta$ the bifurcation angle as in the previous sections.
To investigate properties of a vascular network in the present model, we firstly consider its graphical property as a directed graph.
We will later take into account the cross sections of blood vessels.

\begin{figure}[htbp]
\begin{center}
\includegraphics[scale=0.3]{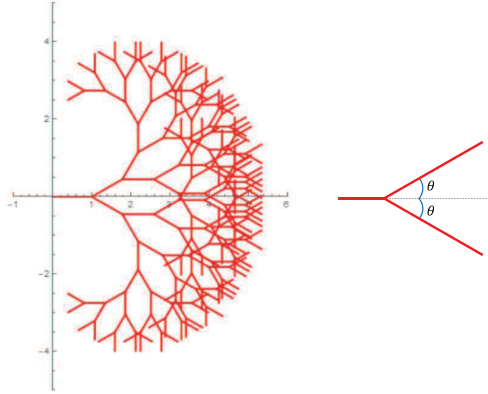}
\caption{Fractal-like tree structure of a model vascular network without reconnection. The ratio of the length of a bifurcated branch to that of its parent branch $r$ is constant for every bifurcation. The bifurcation angle $\theta$ is also the same at every branch point.}
\label{fig:FractalTree}
\end{center}
\end{figure}

We express the positions of a network in a complex $z$-plane.
By taking an appropriate length unit, we suppose that the first blood vessel elongates from the origin $z=0$ along the positive real axis and bifurcates at $z=1$.
Then the next two branch points are given as $1+r\e^{\pm\sqrt{-1} \theta}$.
We denote a set of branch points by $\{z(i,j)\}$ $(i=0,1,2,...;\, j=1,2,..,2^{i-1})$,
where $z(0,1)=1$ and the other points are defined recursively as 
\begin{eqnarray*}
 && z(i + 1,2j-1) - z(i,j) =r^i e^{\sqrt{-1}(k_{i,j} + 1)\theta},\quad  k_{i + 1,2j-1} =k_{i,j}+1, \\
 && z(i + 1,2j) - z(i,j) =r^i e^{ \sqrt{-1}(k_{i,j} - 1)\theta },\quad  k_{i + 1,2j} =k_{i,j}-1
\end{eqnarray*}
with $k_{0,1}=0$.
We also define $z(0,0)=0$. 
When no reconnection is made, for example, in case $r=\frac{9}{10}$ and $\theta=\frac{\pi}{6}$, we obtain a vascular network pattern shown in Fig.~\ref{fig:FractalTree}. 

%
%

\begin{figure}[htbp]
\begin{center}
\includegraphics[scale=0.5]{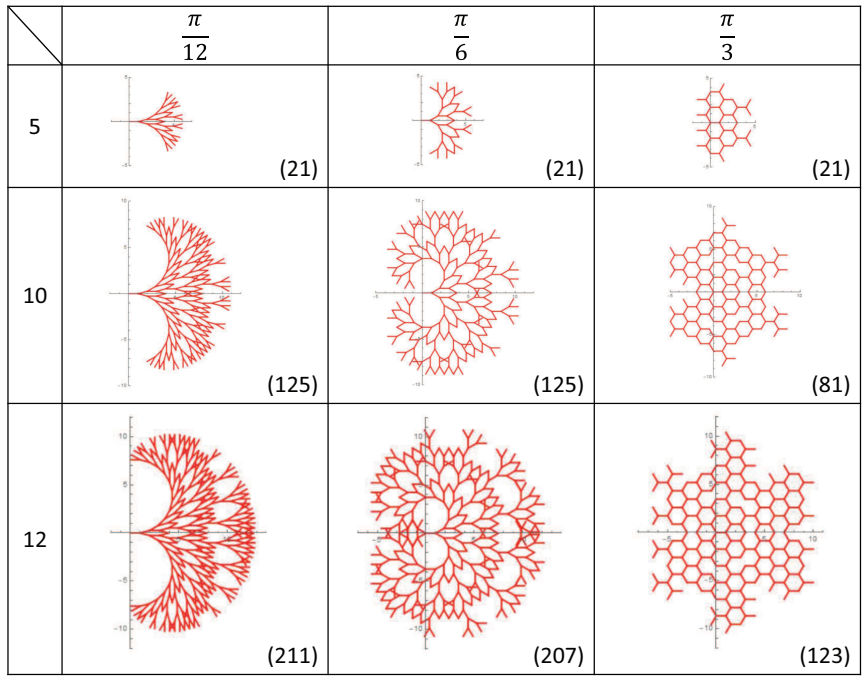}
\caption{Growth of fractal trees ($r=1;\ \theta =\frac{\pi}{3},\ \frac{\pi}{6},\ \frac{\pi}{12}$).
The numbers in parentheses indicate the number of bifurcation points.}
\label{fig:growth}
\end{center}
\end{figure}

\begin{figure}[htbp]
\begin{center}
\includegraphics[scale=0.5]{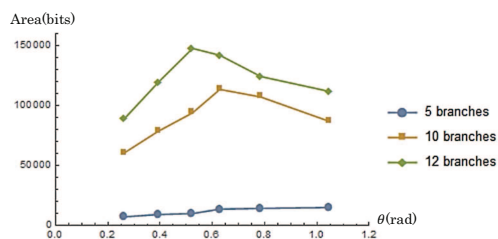}
\caption{Change in total area with increasing number of branch points.}
\label{fig:GrowthCurve}
\end{center}
\end{figure}

Hereafter we take into accaunt the reconnection of branches. Figures~\ref{fig:growth} show example of the fractal-like structure of the patterns with respect to bifurcation angles.
For a sufficiently small angle, the tree structure grows densely in front. 
On the other hand, for a sufficiently large angle, the spread of the tree structure is sparse and branches near its center are short. 
We see that the fractal tree spreads in a well-balanced manner for an intermediate angle. 
The angular variation of the total area covered by the networks in Fig.~\ref{fig:growth} is estimated with image analysis and is shown in Fig.~\ref{fig:GrowthCurve}. 
There is a peak around $\theta \sim \frac{\pi}{6}$, which is an observed angle in the in vitro experiments of angiogenesis\cite{Arima-Nishiyama}. 

%
%

\begin{figure}[htbp]
\begin{center}
\includegraphics[scale=0.4]{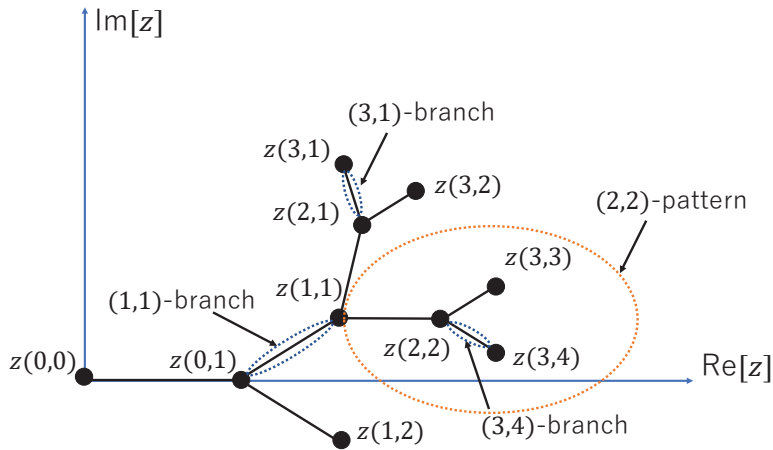}
\caption{A vascular network described in the $z$-plane. The $(1,1)$-branch is the original root branch of the pair of the $(3,1)$-branch and the $(3,4)$-branch.}
\label{fig:point_explain}
\end{center}
\end{figure}

To proceed, let us prepare some terminology. (See Fig.~\ref{fig:point_explain}).  
If a branch A is generated from a branch B after successive bifurcations, we say that B is 
\textit{a root branch} of A.
The root branch common to a pair of two branches that is closest to the pair
is said the \textit{original root branch} of the pair.
Note that, if a branch A is a root of another branch B, the original root branch of the pair A and B is defined as A.
For a branch A, the A-pattern is the partial pattern of a vascular network which consists of the set of all the branches of which A is a root branch.
We also index a branch in a network by a pair of integers $(i,j)$ ($i,\,j  \in \Z_{\ge 0}$, $1 \le j \le 2^i$).
Here $i$ denotes the order of bifurcation, and the integer $j$ distinguishes the $2^i$ branches generated at $i$th bifurcation, such that the branch $(i,j)$ is a root branch of $(i+1,2j-1)$ and $(i+1,2j)$ and the angle between $(i,j)$-branch and $(i+1, 2j-1)$-branch ($(i+1,2j)$-branch) is $\theta$ ($-\theta$). 
In other words, $(i,j)$-branch is the branch the endpoint of which is $z(i,j)$.
Note that  $(0,1)$-branch is the root branch of all the branches.
In a complex $z$-plane, $(0,1)$-branch is expressed as a line from the origin ($z=0$) to the branch point $z(0,1)=1$.

\begin{figure}[htbp]
\begin{center}
\includegraphics[scale=0.5]{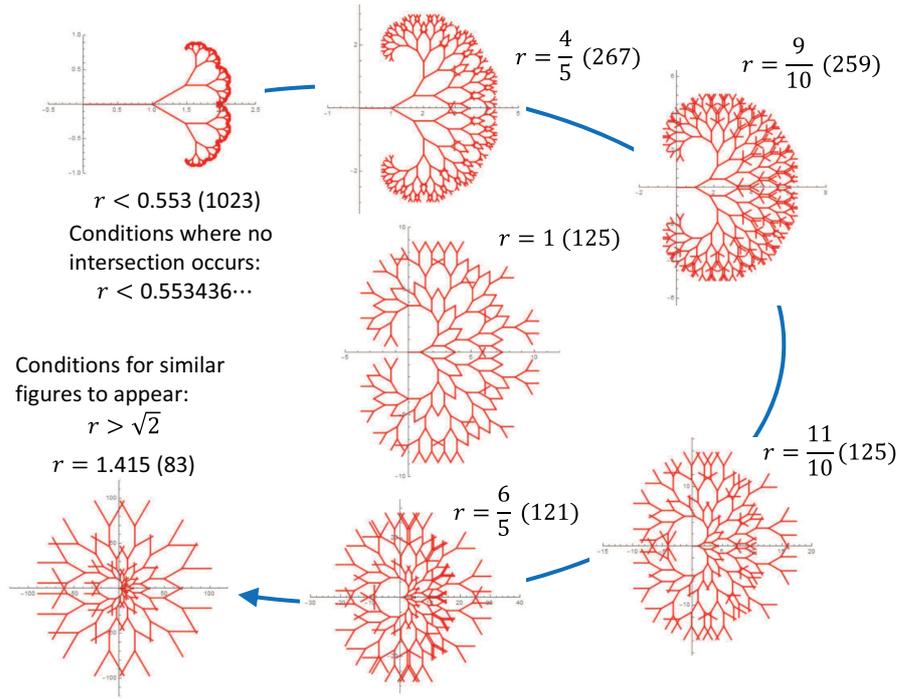}
\caption{Fractal-like tree structure with reconnection effect. The numbers in parentheses indicate the number of branch points.
Note that the scale of each of these figures is different.}
\label{fig:fractal}
\end{center}
\end{figure}

In Fig.~\ref{fig:FractalTree}, we have not considered the reconnection of blood vessels, and the branches intersects indefinitely.
By changing the ratio $r$ with the fixed bifurcation angle $\theta=\frac{\pi}{6}$,  we have vascular network patterns shown in Fig.~\ref{fig:fractal}, where bifurcation is stopped when branches intersect.

\begin{figure}[htbp]
\begin{center}
\includegraphics[scale=0.3]{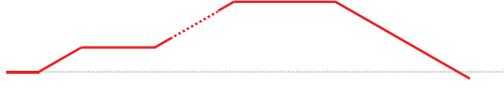}
\caption{Conditions for similar octagonal islands in the pattern to appear. If one of the two branches bifurcated from a branch parallel to the real axis always intersects the real axis, a series of self-similar patterns are created. (See Proposition~\ref{prop0}).}
\label{fig:SimilarFigures}
\end{center}
\end{figure}

An interesting observation is that there appears growing self-similar patterns for $r\ge \sqrt{2}$ (See the lower left figure in Fig.~\ref{fig:fractal}). 
In the network pattern, similar octagons are successively created by reconnection or binding of branches.
In fact, the following proposition holds.
\begin{Proposition}\label{prop0}
For any bifurcation angle $\theta$ ($0<\theta<\frac{\pi}{2}$), a network exhibits self-similar growing pattern with octagons  for $r \ge \sqrt{2}$.
\end{Proposition}
\Proof
Let us consider the staircase path in the pattern which is obtained by connecting the series of branch points:
\[
z(0,1),\ z(1,1),\ z(2,2),\ z(3,3),\ z(4,6),\ \ldots,\ z(i, k_i),\ z(i+1,k_{i+1}),\ \ldots .
\]
Here $k_1=k_2=1$, $k_{i+1}=2k_i-1$ for even $i$, and $k_{i+1}=2k_i$ for odd $i$.
Note that a $(2m,k_{2m})$-branch ($m \in \Z_+$), the endpoint of which is $z(2m, k_{2m})$, is parallel to the real axis.
Then $(2m+1,k_{2m+1}+1)$-branch is the other branch than the $(2m+1, k_{2m+1})$-branch bifurcated from the point $z(2m,k_{2m})$.
Since the network pattern has self-similarity and mirror symmetry about the real axis,
if any of $(2m+1,k_{2m+1}+1)$-branch ($m=1,2,3,...$) intersects the real axis, the similar octagons are created at each bifurcation (Fig.~\ref{fig:SimilarFigures}).
For a given $\theta$, 
\[
\mbox{\rm Im}[z(2m,k_{2m})]=(r+r^3+\cdots +r^{2m-1})\sin \theta=\frac{r(1-r^{2m})\sin \theta}{1-r^2},
\]
and 
\[
\mbox{\rm Im}[z(2m+1,k_{2m+1}+1)-z(2m,k_{2m})]=-r^{2m+1}\sin \theta.
\]
Hence, if for any $m$ ($m=1,2,3,...$),
\[
\frac{r(1-r^{2m})\sin \theta}{1-r^2} -r^{2m+1}\sin \theta \le 0,
\]
we have a self-similar growing pattern like the lower left figure in Fig.~\ref{fig:fractal}.
Thus we find the condition that
\[
2-\frac{1}{r^{2m}} \le r^2\quad (m=1,2,3,...).
\]
Therefore for any angle $\theta$, the pattern has a growing self-similar structure for $r \ge \sqrt{2}$.
\qed\vspace*{5pt}


%
%

Another interesting fact is that there appears no intersection for $r<0.553436\cdots$, and the pattern shows a fractal-like structure. 
In general, there exists a ratio $r_\theta$ for a given $\theta$ such that no intersection takes place for $r \le r_\theta$.
In other words, if $r \le r_\theta$, the network pattern exhibits a fractal-like structure.
Let us determine this ratio as a function of $\theta$ ($0<\theta\le \frac{\pi}{2}$).
The following lemma is intuitively obvious.
\begin{Lemma}\label{lem1}
When the first intersection of branches takes place after a series of bifurcations, the original root branch of the two crossed branches is the $(0,1)$-branch.
\end{Lemma}
\Proof
If the other branch is the original root branch, self-intersection must have occurred before due to the self-similarity of the pattern.
\qed\vspace*{5pt}

From Lemma~\ref{lem1}, the first intersection takes place between $(1,1)$-pattern and $(1,2)$-pattern.
Hence if and only if these patterns do not intersect, there appears no intersection.
Because these patterns are linearly symmetrical to each other with the real axis of symmetry, we have the following Lemma.
\begin{Lemma}\label{lem2}
The network pattern has no intersection point, if and only if $(1,1)$-pattern does not intersect the real axis. 
\end{Lemma}

Since the network is connected, Lemma~\ref{lem2} implies that no intersection takes place if and only if the lower bound of imaginary values of branch points in $(1,1)$-pattern is non-negative.
\begin{Proposition}\label{prop1}
Let $n_{\theta} \in \Z_+$ be the positive integer which satisfies
\begin{equation}\label{sec4_n_theta}
(n_{\theta}-1)\theta<\frac{\pi}{2} \le n_{\theta} \theta,
\end{equation}
then the lower bound of imaginary values of branch points in $(1,1)$-pattern, $\eta_\theta$,  is given by the following formula.
\begin{equation}\label{sec4_lower_value}
\eta_\theta=r\sin \theta  - \sum_{k=1}^{n_{\theta}}r^{k+2}\sin k\theta 
-\frac{r^{n_{\theta}+3}\sin (n_{\theta}-1)\theta +r^{n_{\theta}+4}\sin n_{\theta}\theta  }{1-r^2}.
\end{equation}
\end{Proposition}
%
\Proof
For $i \ge 2$, we denote by $z_i^*$ the branch point whose imaginary value is the lowest among the $i$-th bifurcation points. 
Since $z(1,1)=1+r\e^{\sqrt{-1}\theta}$, $z_2^*=z(2,2)=1+r\e^{\sqrt{-1}\theta}+r^2$,  and $z_3^*=z(3,4)=z_2^*+r^3\e^{-\sqrt{-1}\theta}$.
By estimating similarly, we find that $z_i^*=z(i,2^{i-1})=z_{i-1}^*+r^{i}\e^{-\sqrt{-1}(i-2)\theta}$ for $2 \le i \le n_{\theta}+2$.

\begin{figure}[htbp]
\begin{center}
\includegraphics[scale=0.3]{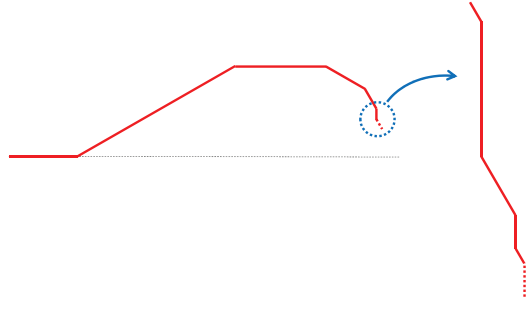}
\caption{Conditions for no intersection of branches. If any branch to which $(1,1)$-branch is a root branch never touches the real axis, there is no intersection of branches anywhere in the pattern. (See Proposition~\ref{prop1}.)}
\label{fig:NoIntersection}
\end{center}
\end{figure}

Let $n_0:=n_{\theta}+2$, then, 
since $\sin (n_{\theta}+1)\theta \ge \sin (n_{\theta}-1)\theta$, 
\[
z_{n_0+1}^*=z_{n_0}^*+r^{n_0}\e^{-\sqrt{-1}(n_{\theta}-1)\theta}.
\]
Thus, we have  
\begin{align*}
z_{n_0+2}^*&=z_{n_0+1}^*+r^{n_0+1}\e^{-\sqrt{-1}n_{\theta}\theta}\\
z_{n_0+3}^*&=z_{n_0+2}^*+r^{n_0+2}\e^{-\sqrt{-1}(n_{\theta}-1)\theta}\\
z_{n_0+4}^*&=z_{n_0+3}^*+r^{n_0+3}\e^{-\sqrt{-1}n_{\theta}\theta}\\
 & \qquad \cdots  \\
 z_{n_0+2m}^*&=z_{n_0+2m-1}^*+r^{n_0+2m-1}\e^{-\sqrt{-1}n_{\theta}\theta}\\
 z_{n_0+2m+1}^*&=z_{n_0+2m}^*+r^{n_0+2m}\e^{-\sqrt{-1}(n_{\theta}-1)\theta}\\
 & \qquad \cdots 
\end{align*}
The path connecting $z_i^*\ (i=1, 2, \ldots)$ is schematically shown in Fig.~\ref{fig:NoIntersection}. 
Because $\eta_{\theta}=\lim_{i \to \infty} \mbox{\rm Im}\left[z_i^* \right]$,
we obtain the expression \eqref{sec4_lower_value}.
\qed

The following Corollary is a direct consequence of Lemma~\ref{lem2} and Proposition~\ref{prop1}. 
\begin{Corollary}
The critical ratio $r_{\theta}$ satisfies the equation
 \begin{equation}\label{sec4_critical_value}
\left. r\sin \theta  - \sum_{k=1}^{n_{\theta}}r^{k+2}\sin k\theta 
-\frac{r^{n_{\theta}+3}\sin (n_{\theta}-1)\theta +r^{n_{\theta}+4}\sin n_{\theta}\theta  }{(1-r)^2}\,\, \right|_{r=r_\theta}=0,
\end{equation}
and no intersection occurs in the network pattern for $r \le r_{\theta}$.
\end{Corollary}

Note that $0<r_{\theta}<1$.
In case $\theta = \frac{\pi}{6}$, $n_\theta=3$ and \eqref{sec4_critical_value} leads
\begin{align*}
0&=r\sin \frac{\pi}{6}-r^3\sin \frac{\pi}{6}-r^4\sin \frac{\pi}{3}-r^5\sin \frac{\pi}{2}
-\frac{r^6\sin \frac{\pi}{3}+r^7\sin \frac{\pi}{2}}{1-r^2}\\
&=\frac{r}{2}-\frac{r^3}{2}-\frac{\frac{\sqrt{3}}{2}r^4+r^5}{1-r^2}\\
&=\frac{r(1-2r^2-\sqrt{3}r^3-r^4)}{2(1-r^2)}.
\end{align*}
Hence we find that $r=r_{\frac{\pi}{6}}$ is given by the root of the equation
\begin{equation}\label{sec4_p6}
1-2r^2-\sqrt{3}r^3-r^4=0.
\end{equation}
Solving \eqref{sec4_p6} numerically, we obtain $r_{\frac{\pi}{6}}=0.553436\cdots$.
Thus we proved that no intersection takes place for $r\le r_{\frac{\pi}{6}}$ as shown in Fig.~\ref{fig:fractal}.

We note that the equation in \eqref{sec4_critical_value} can be rewritten as
\begin{align}
&\frac{r\sin \theta-r^2\sin 2\theta+r^{n_{\theta}+1}\sin(n_{\theta}-1)\theta-r^{n_{\theta}+2}\sin(n_{\theta}-2)\theta}{1-2r\cos \theta +r^2}\nonumber \\
&\qquad \qquad 
-\frac{r^{n_{\theta}+1}\sin (n_{\theta}-1)\theta +r^{n_{\theta}+2}\sin n_{\theta}\theta  }{1-r^2}=0.
\end{align}
In particular, for $n_{\theta}=2$, which implies that $\frac{\pi}{4}\le \theta <\frac{\pi}{2}$, $r=r_{\theta}$ satisfies the cubic equation:
\begin{equation}
1-2r^2-2(\cos \theta )r^3=0.
\end{equation}

%
%
%

\begin{figure}[htbp]
\begin{center}
\includegraphics[scale=0.6]{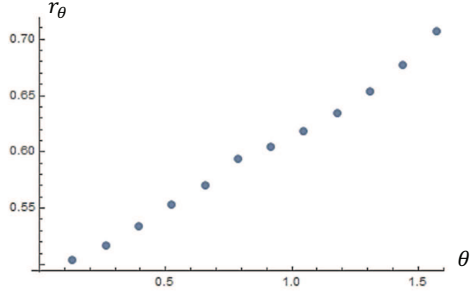}
\caption{Graph of $r_\theta$ where no intersection occurs at each angle $\DIS \theta=\frac{k\pi}{24}\ (k =1,2,\cdots,12)$}
\label{fig:NoInterAng}
\end{center}
\end{figure}

%
%

Next, we determine the Hausdorff dimension (fractal dimension) $D_h$ of these fractal-like tree patterns\cite{Mandelbrot}. 
The total length of the network is calculated as
\[
L_0\left(1+2r+(2r)^2+\ldots\right)=\lim_{n \to \infty}\frac{\left(1-(2r)^n\right)L_0}{1-2r}.
\]
Hence, when $r <\frac{1}{2}$, the total length is finite and its Hausdorff dimension is equal to its topological dimension, that is, $D_h=1$.
While $\frac{1}{2}< r \le r_{\theta}$,  let $\epsilon_n:=\frac{1}{2}r^n$ and $N(n)$ be
the number of the 2-dimensional disks with radius $\epsilon_n$ which is required to cover the network.
We estimate the number $N(n)$ as
\[
\frac{1}{2\epsilon_n}\left(1+2r+...+(2r)^n\right) \le N(n) \le 
\frac{1}{2\epsilon_n}\left(1+2r+...+(2r)^n+ ...+(2r)^{n+n_0}\right),
\]
where $n_0$ is the smallest positive integer which satisfies $r^{n_0}<(1-r)$.
Thus we find that
\[
\frac{(2r)^{n+1}-1}{\epsilon_n(2r-1)}\le N(n) \le \frac{(2r)^{n+n_0+1}-1}{\epsilon_n(2r-1)}.
\]
With inequalities $(2r)^{n+1}-1>(2r)^n$ and $(2r)^{n+n_0+1}-1<(2r)^{n+n_0+1}$,
the following relation holds
\[
\frac{2}{2r-1}2^n<N(n)<\frac{2^{n_0+1}r^{n_0}}{2r-1}2^n.
\]
The Hausdorff dimension, which is equal to the box-counting dimension, is given by
\[
D_h=\mbox{inf}\left\{d \ge 0\, \big|\, \lim_{n \to \infty}  N(n)\epsilon_n^d <+\infty      \right\},
\]
which leads to the relation  $2(r)^{D_h}=1$.       
Therefore we obtain
\begin{Proposition}
\begin{equation}\label{sec4_H_dim}
D_h=-\frac{\log 2}{\log r}. 
\end{equation}
\end{Proposition}
Since $r$ satisfies the relation $\frac{1}{2} \le r \le r_{\theta} \le \frac{1}{\sqrt{2}}$
for a non-intersecting network, we find that $1 \le D_h \le 2$.

Fractal dimension is also estimated from the self-similarity of the pattern.
If we enlarge the pattern by a factor of $\frac{1}{r}$, we have two same patterns and one line segment. Since the measure of the line segment is negligible, the fractal dimension (similarity dimension) $D_f$ of the pattern is evaluated by
\[
\left( \frac{1}{r} \right)^{D_f}=2.
\] 
Thus we have that $D_f=D_h= -\frac{\log 2}{\log r}$ again.

So far we have not taken into account the fact that the network consists of vessels with finite cross sections.
In the present model, the density of ECs in the vessels is assumed to be equal, and each branch is characterized by the number of ECs per unit length.
The volume of the pattern up to the branches generated the $n$th order bifurcation  
 is given as
\begin{align*}
V_n&=L_0s_0+2L_1s_1+2^2L_2s_2+...+2^nL_ns_n\\
&=L_0s_0\left( 1+2^{1-\frac{1}{\mu}}r+(2^{1-\frac{1}{\mu}}r)^2+...+(2^{1-\frac{1}{\mu}}r)^2\right)\\
&=\frac{L_0s_0\left(1-(r^*)^{n+1}\right)}{1-r^*},
\end{align*}
where $r^*:=2^{1-\frac{1}{\mu}}r$.
Since $\mu \sim \frac{3}{2}$ and $r <\frac{1}{\sqrt{2}}$ for a non-intersecting network, $r^*<1$ and the total volume $V_\infty$ of the pattern is
\[
V_\infty=\lim_{n \to \infty}V_n=\frac{L_0s_0}{1-r^*}.
\] 
Then total number of ECs in a pattern is equal to $\rho_b V_\infty =:N_{tot}$. 
Just as the discussion about  the fractal nature of DLA, coastline shape, images of retina and so on,  the present fractal-like structure is an approximation of the actual blood vessel network, and the branches of which lengths are less than the size of an EC is meaning-less.
So it is not surprising that the total number of ECs is finite though the total length of the pattern is infinite.
The $(i,j)$-branch has $\sigma_i:=\rho_b s_{i-1}$ ECs per unit length in average, and the ratio $\mu_i:=\frac{\sigma_i}{N_{tot}}$ gives a measure of the $(i,j)$-branch, that is, 
for any part of the network pattern, $g$,  the ratio of the number of ECs in $g$ to the total number of ECs is given as 
\[
p(g):=\int_{x \in g} d\mu(x),
\]
where, with ordinary Lebesgue measure $dx$ of a line, 
\begin{equation}\label{sect4_measure}
d\mu(x):=\mu_i dx \qquad (x \in \mbox{$(i,j)$-branch}).
\end{equation}
Once the measure $d\mu(x)$ is defined, we can try so called multifractal analysis\cite{HJKPS,SMA}. 
Hereafter we set $L_0=1$ for simplicity.
Let us consider a large integer $n$ ($n \gg 1$) and define $l_n:=r^n$, which is the length of the branch created at the $n$th bifurcation.
We denote by $G_n$ the partial pattern that is created at $n$th bifurcation, and divide it into a set of segments with length $l_n$.
Then the partition function $Z_n(q)$ is defined as
\begin{equation}
Z_n(q):=\sum_{s \in \{\mbox{\scriptsize all segments in $G_n$}\}}  p(s)^q. 
\end{equation}
The original branch ($(0,1)$-branch) is divided into $\sim \frac{1}{l_n}=r^{-n}$ pieces of segments, and each of them has a measure:
\[
p(s)\Big|_{s\in \{\mbox{\scriptsize segments in $(0,1)$-branch}\}}=p_0:=l_n \mu_0.
\] 
The $(1, j)$-branch  $(j=1,2)$ is divided into $\sim \frac{r}{l_n}=r^{-n+1}$ pieces of segments and each of them has a measure $p_1:=l_n \mu_1$.
Similarly, the $(k,j)$-branch is divided into  $\sim \frac{r^{k}}{l_n}=r^{-n+k}$ pieces of segments and each of them has a measure $p_k:=l_n \mu_k$.
Thus we have
\[
\mu_k=\frac{\sigma_k}{N_{tot}}=\frac{\rho_b s_{k}}{\rho_bs_0}(1-r^*)=2^{-\frac{k}{\mu}}(1-r^*),
\]
and
\[
p_k=r^n 2^{-\frac{k}{\mu}}(1-r^*).
\]
Therefore we have
\begin{align*}
Z_n(q)&=\sum_{k=0}^n 2^{k}r^{-n+k} p_k^q \\
&=\sum_{k=0}^n2^{k}r^{-n+k+nq}2^{-\frac{qk}{\mu}}(1-r^*)^q\\
&=(1-r^*)^{q-1}\left(1-(r^*)^{n+1}\right)r^{n(q-1)}\\
&\sim \left\{
\begin{array}{cl}
l_n^{q-1}&\  \left(q>\mu\left(1+ \frac{\log r}{\log 2}\right)\right) \\
l_n^{q+\left(1-\frac{q}{\mu} \right)\frac{\log 2}{\log r}} &\ \left(q< \mu\left(1+ \frac{\log r}{\log 2}\right)\right)
\end{array} .
\right.
\end{align*}
Since multifractal dimension $D_q$ is given by
\[
Z_n(q) \sim l_n^{(q-1)D_q},
\]
we have the following Proposition.
\begin{Proposition}
With the measures $\mu(x)$ defined in \eqref{sect4_measure}, the multifractal dimension $D_q$ is determined as 
\begin{equation}
D_q=\left\{ 
\begin{array}{cl}
1&\  \left(q>\mu\left(1+ \frac{\log r}{\log 2}\right)\right)  \\
\frac{\mu q \log r +(\mu-q)\log 2}{\mu(q-1)\log r}&\   \left(q<\mu\left(1+ \frac{\log r}{\log 2}\right)\right)  
\end{array}
\right..
\end{equation}
\end{Proposition}
$D_0=-\frac{\log2}{\log r}$ is the Hausdorff dimension, which is, as expected,  equal to \eqref{sec4_H_dim}.
In the present situation, $\mu \sim \frac{3}{2}$, $\frac{1}{2}<r<\frac{1}{\sqrt{2}}$, and $0<\mu\left(1+ \frac{\log r}{\log 2}\right)<1$.
Hence both the information dimension $D_1$ and the correlation dimension $D_2$ are equal to $1$.

%
%

So far we have considered fractal aspects of vascular networks for idealized models in which the branching ratio $r$ is the same at every bifurcation.
Let us examine the same analysis to the network patterns with reconnection in the previous section.
We consider the cases shown in Fig.~\ref{fig:distribution}.
The bifurcation angle is fixed to $\frac{\pi}{6}$, Murray's index $\mu=\frac{3}{2}$ and the variance of the parameters $\lambda_k^{(i)}$ is set to $\sigma=0.1$.
We calculated the multifractal dimension $D_q$ for the patterns with 13 and 19 bifurcations.
The measure $d\mu(x)$ for the network is taken to be the same as above.
The specific steps to obtain $D_q$ of a given pattern is as follows;
\begin{enumerate}
\item Calculate all the branch lengths $L_k^{(i)}$ according to the differential equations and reconnection procedure.
\item For the size of the pattern $A$, we rescale $L_k^{(i)}\, \rightarrow \, L_k^{(i)}/A$. Here, $A$ is approximately given as $A = \sum_{k=0}^{k_{max}}L_k^{(1)}$. (Note that summation is taken over only $k$.) 
\item Let $N_{tot}:=\sum_{k=0}^{k_{max}} \left[\sum_i L_k^{(i)}\right] (2^{-k/\mu}) $,
$l:=\min\left[ L_k^{(i)}\right]$, and $s_k:=2^{-k/\mu}$.
\item For $q\in \mathbb{R}$, we define 
\[
Z(q):=\sum_{k=0}^{k_{max}} \left[\sum_i \frac{L_k^{(i)}}{l}\right] \left(\frac{s_k l}{N_{tot}}\right)^q,
\]
and 
\[
D_q=\frac{\log Z(q)}{(q-1) \log l}.
\]
\end{enumerate}

\begin{figure}[hbtp]
\begin{center}
\includegraphics[scale=0.5]{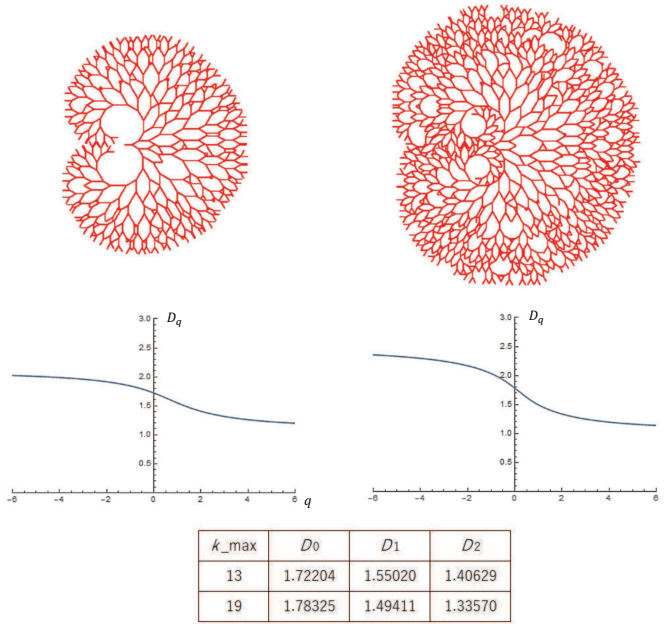}
\caption{Vascular networks with reconnection and their multifractal dimension $D_q$.  Bifurcation angle is $\frac{\pi}{6}$ and variance $\sigma=0.1$. Branches are counted up to $13$th bifurcation and $19$th bifurcation respectively.}
\label{fig:multi-fractal_Re}
\end{center}
\end{figure}

The results are shown in Fig.~\ref{fig:multi-fractal_Re}.
We do not have any experimental data on angiogenesis to compare this result with, but, for human retinal vessels, it was reported that $D_0 \sim 1.65$, $D_1 \sim 1.6$ and $D_2 \sim 1.55$ in Ref.~\cite{StosicStosic}.
A retinal vascular network is fairly different from the vascular networks in angiogenesis, because, for example, it is constructed so as not to interfere with optic nerves\cite{OKYKTMMSSEK},  but its qualitative trends of multifractal analysis are consistent to our results in Fig.~\ref{fig:multi-fractal_Re}.


\section{Concluding remarks}
\label{Chap:conclusion}

We have investigated the simultaneous differential equation model for angiogenesis
described by \eqref{sec2_difeq_a} -- \eqref{sec2_difeq_d}.
The present model is a slight modification of the mathematical model given in Ref.~\cite{MMYKT} which was proposed as a continuous approximation of the dynamics of  ECs in angiogenesis.
We gave several cases where the exact solutions can be obtained analytically.  
We have also proved the conditions under which self-similarity appears in the patterns. 
In order to consider the vascular network to which blood vessels are connected like the blood vessels of a retina, we analysed the vascular networks with the effect of reconnection, and obtain the distribution of area of islands in the vascular network.   
As a result, we found that pattern formation changes significantly depending on the branch length and bifurcation angle of the vascular network. 
In particular, a pattern with bifurcation angle around $\frac{\pi}{6}$, which is close to that often observed in in vitro experiments, has an efficient feature of covering a large area with small total length of blood vessels. 
Furthermore, by cluster analysis, the distribution of the islands is shown to exhibit power-law behaviour for the patterns with bifurcation angle around $\frac{\pi}{6}$.
Such behaviour was also reported in retinal vasculature in mice\cite{IJTYM}. 
To characterise the resulting vascular pattern, multifractal analysis was considered analytically in the case of idealised self-similar vascular networks and numerically in the case of the influence of reconnection.

In the future, we would like to improve the present mathematical model by adjusting functions and parameters based on the experimental results, and analyse the actual vascular network in angiogenesis using the present method.

%
%
%
%
%
%
%
%

\begin{acknowledgements}
The authors would like to thank Prof. Hiroki Kurihara and Dr. Kazuo Tonami for helpful discussions about angiogenesis. They also would like to thank Dr. Tatsuya Hayashi and Mr. Kazuma Sakai for useful discussions on mathematical modelling. TT is grateful for financial support to Arithmer Inc. .    
\end{acknowledgements}

%
%



\end{document}